\documentclass[twocolumn, 12pt]{aa}
\pdfoutput=1

\usepackage[english]{babel}
\usepackage{multicol}
\usepackage{amsmath}
\usepackage{amsfonts}
\usepackage{dsfont}
\usepackage{amsxtra}
\usepackage[draft]{hyperref}
\usepackage{amssymb}
\usepackage{upgreek} 
\usepackage{changes}
\usepackage{multirow}
\usepackage{url}
\usepackage{float}

\usepackage{graphicx,epsfig}
\usepackage{dcolumn}

\usepackage{xspace} 

\usepackage{color}
\usepackage{xcolor}

\usepackage[switch]{lineno}
\usepackage{lipsum}

\newcommand{\be}{\begin{equation}}
\newcommand{\ee}{\end{equation}}
\newcommand{\bea}{\begin{eqnarray}}
\newcommand{\eea}{\end{eqnarray}}
\newcommand{\beaa}{\begin{eqnarray*}}
\newcommand{\eeaa}{\end{eqnarray*}}
\newcommand{\ba}{\begin{array}}
\newcommand{\ea}{\end{array}}
\newcommand{\bi}{\begin{itemize}}
\newcommand{\ei}{\end{itemize}}
\newcommand{\ben}{\begin{enumerate}}
\newcommand{\een}{\end{enumerate}}

\newcommand{\lb}{\label}

\newcommand{\Fermi}{\textit{Fermi}\xspace}

\definecolor{darkgreen}{rgb}{0.0, 0.7, 0.0}

\begin{document}

   \title{Energy dependent morphology of the pulsar wind nebula HESS\,J1825$-$137 with \textit{Fermi}-LAT}


   \author{G. Principe \thanks{\email{giacomo.principe@inaf.it}}
          \inst{1,2}
          \and
		 A.M.W. Mitchell 
          \inst{3}
          \and   
          S. Caroff
          \inst{4}
          \and
          J. A. Hinton 
          \inst{5}
		  \and
          R.D. Parsons 
          \inst{5}
          \and
          S. Funk 
          \inst{2}          
          }

   \institute{
             {INAF - Istituto di Radioastronomia, Bologna, Italy}
             \and
            {Erlangen Centre for Astroparticle Physics, Erlangen, Germany}
            \and
               {Physik-Institut of the University of Zurich, Zurich, Switzerland}
            \and
               {Sorbonne Université, Université Paris Diderot, Sorbonne Paris Cité, CNRS/IN2P3, Laboratoire de Physique Nucléaire et de Hautes Energies, LPNHE, 4 Place Jussieu, F-75252 Paris, France}
            \and
               {Max-Plank-Institut f\"ur Kernphysik, Heidelberg, Germany}
             }

   \date{Received May 9, 2020; accepted June 15, 2020}


  \abstract
   {} 
   {Taking advantage of more than 11 years of \textit{Fermi}-LAT data, we perform a new and deep analysis of the pulsar wind nebula (PWN) HESS J1825–137. Combining this analysis with recent H.E.S.S. results we investigate and constrain the particle transport mechanisms at work inside the source as well as the system evolution.}
   {The PWN is studied using 11.6 years of \textit{Fermi}-LAT data between 1\,GeV and 1\,TeV. In particular, we present the results of the spectral analysis and the first energy-resolved morphological study of the PWN HESS\,J1825$-$137 at GeV energies, which provide new insights into the $\gamma$-ray characteristics of the nebula.}
   {An optimised analysis of the source returns an extended emission region larger than 2$^{\circ}$, corresponding to an intrinsic size of about 150\,pc, making HESS\,J1825$-$137 the most extended $\gamma$-ray PWN currently known. The nebula presents a strong energy dependent morphology within the GeV range, moving from a radius of $\sim1.4^\circ$ below 10\,GeV to a radius of $\sim$0.8$^\circ$ above 100\,GeV, with a shift in the centroid location.}
   {Thanks to the large extension and peculiar energy-dependent morphology, it is possible to constrain the particle transport mechanisms inside the PWN HESS\,J1825$-$137. Using the variation of the source extension and position, as well as the constraints on the particle transport mechanisms, we present a scheme for the possible evolution of the system. 
   Finally, we provide an estimate of the electron energy density and we discuss its nature in the PWN and TeV halo-like scenario.
   }
   {}


\maketitle

\section{Introduction}
\lb{sec:intro}

\begin{figure*}[h]
\centering
\includegraphics[width=14cm]{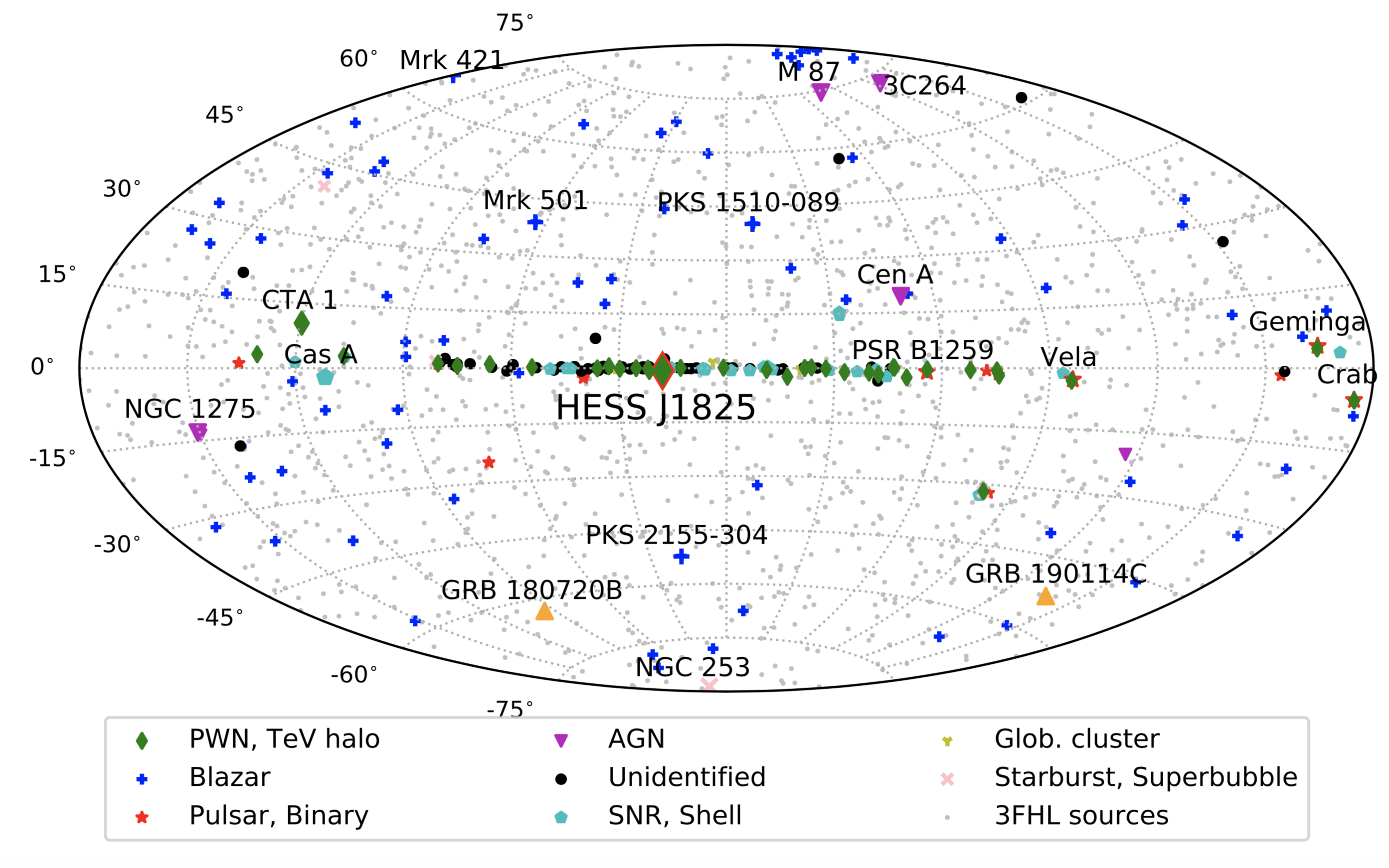}
\caption{\small \label{fig:tevcat_map}
Sky map, in Galactic coordinates and Mollweide projection, showing the sources in the TevCat catalogue
(http://tevcat.uchicago.edu, version April 2019) classified by their most likely association. All the 3FHL sources \citep{2017ApJS..232...18A} are also plotted, with grey points, for a comparison.}
\end{figure*}

Most of the spin-down luminosity of young and very energetic pulsars is carried away in a magnetised wind of charged particles. The confinement of this particle wind outflow, which is predominantly composed of electron-positron pairs, leads to the development of a pulsar wind nebula (PWN). PWNe form when the particle wind collides with its surroundings, especially the slowly-expanding ejecta of the progenitor supernova, and forms a termination shock \citep{2006ARA&A..44...17G}. 
Evolved PWNe are ideal candidates for investigating particle transport mechanisms and electron cooling inside celestial object, due to their large $\gamma$-ray extension and possible energy-dependent morphology, which can provide spatially resolved spectra under certain circumstances. 
PWNe have been observed to emit photons up to TeV energies and, with more than 35 detections, they dominate the population of TeV gamma-ray sources in the Galactic plane (see Fig. \ref{fig:tevcat_map}).
Despite deep observations of several PWNe, many open questions still remain; in particular, the mechanism by which the particles are accelerated at the termination shock is not yet understood \citep{2017hsn..book.2159S}.

\noindent Among such objects, HESS\,J1825$-$137 is the largest and one of the most TeV efficient $\gamma$-ray PWNe currently known \citep{2018A&A...612A...2H,2019A&A...621A.116H}. It is powered by a young and very energetic pulsar PSR\,J1826$-$1334 (also known as PSR\,B1823$-$13), which was discovered by \citet{1992MNRAS.254..177C}.
The pulsar has characteristics very similar to the Vela pulsar: it has a spin period of 101.48\,ms, a characteristic age of 21\,kyr, a spin-down energy of $2.8 \times 10^{38}$ erg s$^{-1}$, and is at a distance of 3.9$\pm$0.4 kpc \citep{2005AJ....129.1993M,2002astro.ph..7156C}. 

The first detection of the extended nebula was made by \citet{1996ApJ...466..938F} using X-ray observations with \textit{ROSAT}, which observed a compact nebula of $\sim20''$ radius around the pulsar.
Subsequent X-ray observations made by \textit{Chandra} and \textit{Suzaku} revealed its asymmetric morphology and an extended emission up to $15'$ (17 pc) \citep{2009PASJ...61S.189U}. 
The discovery of the energy-dependent morphology of HESS\,J1825$-$137 at TeV energies \citep{2006A&A...460..365A} provided important proof that the emission is dominated by `relic' electrons from the earlier epochs of the nebula in which the pulsar was spinning down more rapidly, therefore releasing more energy into the system. 
More recently, \citet{2019ICRC...36..734M} has shown that HESS\,J1825$-$137, together with eHWCJ1908+063 and eHWCJ2019+368, are the only three sources detected above 100 TeV by the HAWC experiment.

In the GeV regime the source was detected first in 2011 by the Large Area Telescope \citep[LAT,][]{2009ApJ...697.1071A}, on board the \textit{Fermi Gamma-ray Space Telescope}, whilst at MeV energies the source is not significantly detected \citep{2018A&A...618A..22P}. Previous LAT analyses of the PWN HESS\,J1825$-$137 have been performed using 20 months of Pass 6 data in the 1 -- 100\,GeV energy band \citep{2011ApJ...738...42G} and subsequently six years of Pass 8 data in the 10\,GeV -- 1\,TeV energy band \citep{2017ApJ...843..139A}. Taking advantage of more than 11 years of \textit{Fermi}-LAT data now available, we have performed an analysis of the energy-dependent morphology and of the spectral parameters of the source in the energy range between 1\,GeV and 1\,TeV.

\section{\textit{Fermi}-LAT data and analysis}
\lb{sec:data}

The LAT is a $\gamma$-ray telescope that detects photons by conversion into electron-positron pairs and has an operational energy range from 20\,MeV to 2\,TeV. It is comprised of a high-resolution converter tracker (for direction measurement of the incident $\gamma$-rays), a CsI(Tl) crystal calorimeter (for energy measurement), and an anti-coincidence detector to identify the background of charged particles \citep{2009ApJ...697.1071A}.

\subsection{Data selection}
For the LAT analysis of HESS\,J1825$-$137 we used 11.6 years of Pass 8 (P8R3) Source class events \citep{2013arXiv1303.3514A, 2018arXiv181011394B} collected between August 4, 2008, and March 20, 2020 (\textit{Fermi} Mission Elapsed Time 239587201 s  -- 606355205 s) in the energy range between 1\,GeV and 1\,TeV. The data were taken in a region of interest (ROI) of radius 15$^{\circ}$ and centred on the source position given in \citet{2017ApJ...843..139A}.
In the following we often use the term `PWN' to refer to the source HESS\,J1825-137.
We created sky maps with a pixel size of 0.1$^\circ$. In order to eliminate most of the contamination from secondary $\gamma$-rays from the Earth’s limb, we excluded $\gamma$-rays with zenith angle larger than 105$^{\circ}$ \citep{2009PhRvD..80l2004A}. We used the 
P8R3\_Source\_V2 instrument response functions (IRFs).

\subsection{Region modelling}
\label{roi_modelling}
The model used to describe the sky includes all point-like and extended LAT sources, within 20$^{\circ}$ degrees of the source position, listed in the fourth \textit{Fermi}-LAT source catalogue \citep[4FGL,][]{2020ApJS..247...33A}, as well as the Galactic diffuse and isotropic emission. 
We modelled the Galactic diffuse emission using two different templates, repeating the analysis for each, in order to study the systematic uncertainty due to the choice of the diffuse model. The first template used (D1) was the Galactic and isotropic diffuse templates\footnote{https://fermi.gsfc.nasa.gov/ssc/data/access/lat/ \\BackgroundModels.html} (labelled in our analysis as `D1') used in the 4FGL model.
For a crosscheck, we used as a second the template the diffuse model\footnote{https://www-glast.stanford.edu/pub\_data/1220/} derived in \citet[][labelled in our analysis as `D2']{2017ApJ...840...43A}
which was especially developed for analysis of extended emission near the Galactic centre. 
The residual background and extragalactic radiation were described by a single isotropic component with the spectral shape in the tabulated model iso\_P8R3\_SOURCE\_V2\_v01.txt.
For the analysis results presented here, we used the first template for the diffuse model unless specified otherwise.

\subsection{Analysis procedure}
\label{sec:analysis_procedure}
The analysis was performed with Fermipy\footnote{http://fermipy.readthedocs.io/en/latest/} \citep[version 0.17.4,][]{2017arXiv170709551W}, a python package that facilitates analysis of data from the LAT with the \textit{Fermi} Science Tools, of which the version 11-07-00 was used. In our analysis we applied the correction for the energy dispersion, as implemented in Fermipy, disabling it for the isotropic model. 

In addition to the study of the general characteristics of the source (localisation, averaged extension and spectra) using the complete energy range between 1\,GeV and 1\,TeV, in this work, we studied also the PWN's morphology in smaller energy bands. The description of the model used to describe the data is presented in Sect. \ref{roi_modelling}. We modelled the PWN using a 2D-Gaussian model for the spatial template and a LogParabola spectral model
\begin{equation} \label{eq:logparabola}
  \left(\dfrac{dN}{dE} = N_{0}\left(\frac{E}{E_{0}}\right)^{- [\alpha + \beta log(E/E_{0})]}\right) ; 
\end{equation}
\noindent as used in the 4FGL, as well as in \citet{2017ApJ...843..139A,2019A&A...621A.116H}.

After a preliminary optimisation (fermipy.optimize) and fit (fermipy.fit) of the parameters of sources included in the model, we investigated (using fermipy.find\_src) the possible presence of additional faint sources, not in the 4FGL catalogue, and we found three new candidate sources that we added to our model. The best-fit positions of these new sources are R.A., decl. = (18$^{h}$13$^{m}$38$^{s}$, -17$^{\circ}$49$^{'}$47$^{''}$),(18$^{h}$18$^{m}$28$^{s}$, -9$^{\circ}$56$^{'}$23$^{''}$), and
(18$^{h}$29$^{m}$34$^{s}$, -16$^{\circ}$14$^{'}$23$^{''}$), with 95\% confidence-level uncertainty R$_{95}=6^{'}44^{''}$. 
We verified also the possible influence of PSR\,J1826-1256 (4FGL\,J1826.1-1256), a bright \textit{Fermi}-LAT source with significant emission at the energies considered in this analysis and that lies at 1$^{\circ}$ from the PWN centre. The gamma-ray steady emission from a point source spatially coincident to the pulsar position is significantly detected in our analysis (TS=5285) with a high flux ($F_{E>1\, \textrm{ GeV}}=(5.55\pm0.06 \times 10^{-8}$ ph cm$^{-2}$ s$^{-1}$).
We verified the possible influence of the pulsar by performing an analysis using only off-phase data, characterised in the forthcoming third \textit{Fermi}-LAT pulsar catalogue using gamma-ray pulsar timing as detailed in \citet{LATtiming}. We found that the morphology and spectral results are compatible within the errors to the results obtained using the full phase data. Furthermore we did not see in our analysis (1 GeV - 1 TeV) any extended residual around the PSR location that would be possibly associated to the VHE extended source HESS\,J1826-130 seen by \citep{2017AIPC.1792d0024A}. The emission of HESS\,J1826-130 is, however, detected by H.E.S.S. only at energies above 2 TeV, which are not covered by \textit{Fermi}-LAT.

Following this, in order to improve the spatial and spectral modelling of HESS\,J1825$-$137, we performed a localisation (fermipy.localize), extension (fermipy.extension) and spectral (fermipy.sed) analysis of the PWN for the entire energy range.
For the spectra derivation using the entire energy range, we divided all the photons between 1\,GeV and 1\,TeV in 24 energy bins (8 per decade) logarithmically distributed. We left all spectral parameters of the diffuse background as well as those of the sources within a 3$^{\circ}$ radius from our target free to vary. For the sources in a radius between 3$^{\circ}$ and 6$^{\circ}$ away, only the normalisation was fitted, while we fixed the spectral parameters of all the sources within the ROI at larger angular distances from our target.

Subsequently, we studied the energy-dependent morphology of the PWN, for which we grouped the photons into five energy bands: four logarithmically spaced bands between 1 and 100\,GeV and one band between 100\,GeV and 1\,TeV.
In the analysis of the extension variation with energy (see Section \ref{ext_2DGaussian}), we fixed the spectra of HESS\,J1825$-$137 (except the normalisation), as well as those of all the other nearby sources, using the resulting model from the full energy range spectral analysis. Only the normalisation of the diffuse and isotropic backgrounds were re-fitted during the analysis of the energy dependent morphology.
Finally, we generated the SED by doing a spaced spectral analysis in each energy band with their corresponding spatial model.

\subsection{Systematic uncertainties due to the emission models used}
\label{sec:systematic_uncertainties}
For the specific goal of a further investigation of the bias introduced by the choice of the diffuse model, we also performed the analysis with other two different models: a template used in \citet[][z=4, ts=150]{2016ApJS..224....8A} and the LAT diffuse emission ring-hybrid model gll\_iem\_v06.fits \citep{2016ApJS..223...26A}, and we compared the results of the four different models used in order to have an estimate of the systematic error due to the diffuse model. Comparing the results of the morphology for the entire energy range 1\,GeV -- 1\,TeV, we found a deviation of the source localisation of $\sim 0.08^{\circ}$, which is similar to the LAT PSF size at 10 GeV (i.e. $\sim$ 0.1$^{\circ}$), and we considered it as an estimate of the systematic error due to the diffuse model. This deviation is more pronounced, $\sim 0.17^{\circ}$ at low energy (E$<10$ GeV), where the diffuse emission is much brighter than at high energy. 
Similarly, during preliminary studies of this source \citep{PrincipeORCiD:2019iiy, 2019ICRC...36..595P}, we performed the analysis modelling the ROI using the sources contained in the FL8Y\footnote{A preliminary version of the 4FGL catalogue,\\ https://fermi.gsfc.nasa.gov/ssc/data/access/\\
lat/fl8y/gll\_psc\_8year\_v5.fit.} which has been derived using standard LAT diffuse emission ring-hybrid model, and we compared the results.
The resulting extension and spectral parameters obtained with these different background models are compatible within the uncertainties.

\section{\textit{Fermi}-LAT results for the entire GeV domain}
In this section we present the results of the spectral and morphological analysis of the PWN HESS\,J1825$-$137 using 11.6 years of \textit{Fermi}-LAT data between 1\,GeV and 1\,TeV.
We also compare the morphological results with previous LAT analyses and combine the spectra obtained in this work with that published in \citet{2019A&A...621A.116H}.

\subsection{Localisation and extension analysis for whole energy range}
\label{sect:avg_loc_ext}
\begin{figure}[h]
\hspace*{-0.4cm}
\centering
\includegraphics[width=1.1\columnwidth, trim=10mm 0mm 10mm 10mm,clip=true]{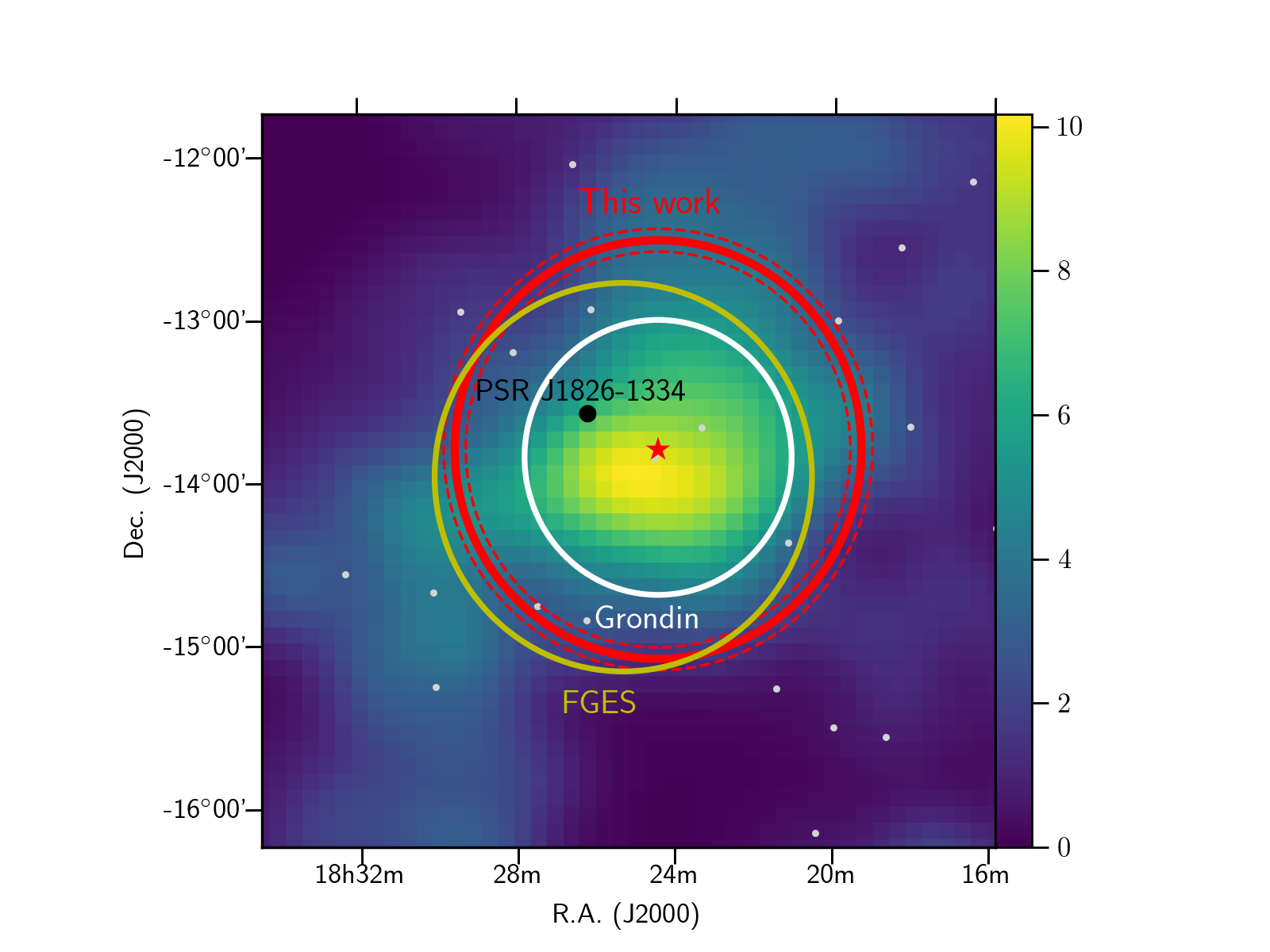}
\caption{\small \label{counts_map} TS map (in sigma units), in celestial coordinates, of the region around the PWN HESS\,J1825$-$137 in the energy range between 1\,GeV and 1\,TeV. The red circle and star indicate the 2D-Gaussian extension and centroid fit obtained in this work. The red dashed circles mark the uncertainty on the 2D-Gaussian extension. The green and white circles correspond to the extension obtained in the FGES catalogue \citep{2017ApJ...843..139A} and in \citet{2011ApJ...738...42G}, respectively. The black point indicates the position of PSR\,J1826-1334, the pulsar which is believed to power the nebula. The 4FGL sources, as well as the three candidate sources added in the model, are represented with light grey points.}
\end{figure}

We performed the localisation and extension analysis using the entire energy range 1\,GeV -- 1\,TeV. For the determination of the extension, we investigated the source radius starting from a value of 0.05$^\circ$ (similar to the best PSF reached), which corresponds to the case of a point source, up to a maximum radius of 2.5$^{\circ}$, in 41 linearly separated steps. 
Fig. \ref{counts_map} shows the test statistic\footnote{The test statistic is the logarithmic ratio of the likelihood of a source being at a given position in a grid to the likelihood of the model without the source, TS=2log($\frac{likelihood_{src}}{likelihood_{null}}$) \citep{1996ApJ...461..396M}.} (TS) map of the region around the PWN for the energy range between 1\,GeV and 1\,TeV. The
TS was evaluated by placing a point source at the centre of each pixel, Galactic diffuse emission and nearby sources being included in the background model. In the map, the size obtained for the PWN in this work is compared with the radius obtained in the previous analysis with \textit{Fermi}-LAT data. 

The best-fitting radius obtained in this work is larger than those obtained in previous works \citep[FGES catalogue, ][]{ 2017ApJ...843..139A,2011ApJ...738...42G}. This is probably connected to the larger amount of data, down to 1\,GeV, which allowed the more extended emission below 10\,GeV to be resolved (see Sect. \ref{ext_2DGaussian}).
Although the pulsar PSR\,J1826$-$1334 is detected in radio and X-rays \citep{2019A&A...623A.115D}, its emission is not significantly detected by \textit{Fermi}-LAT yet. A search for the pulsar was performed looking at a possible steady emission (no pulsation search has been done) and no significant point source emission from this position was found between 1\,GeV and 1\,TeV, only extended emission from the nebula is detected.

Table \ref{table_loc} reports the results of the localisation and extension analysis performed in the energy range between 1\,GeV and 1\,TeV using a 2D-Gaussian model as spatial template for the sources. The extension result reported here corresponds to the 68\% containment radius. 

\begin{table}[h]
\caption{\small \label{table_loc}
Localisation and extension results. The PWN position corresponds to RA(J2000): $276.11^\circ \pm 0.04^\circ$ and Dec.(J2000): $-13.80^\circ \pm 0.04^\circ$. The systematic uncertainty on the localisation and extension estimates due to the diffuse model is 0.08$^{\circ}$, as discussed in Sect. \ref{sec:systematic_uncertainties}}
\small
\centering
\begin{tabular}{cc}
Parameter & Value \\
\hline
\hline
RA & 18h24m26s$\pm$3m8s  \\ 
DEC & $-13^{\circ}47'56''\pm2'14''$  \\
Extension $R_{68\%}$ & $1.30^\circ \pm 0.06^\circ $\\
TS & 1331\\
TS$_{\mathrm{ext}}$ & 1040\\
\hline
\end{tabular}
\end{table}

\noindent The resulting centre position of the PWN is shifted by about 0.48$^\circ$ from PSR\,J1826$-$1334, the pulsar associated with the nebula. This asymmetry of the nebula extension with respect to the PWN position is related to the pulsar proper motion (discussed also in Sect. \ref{system_evolution}) as well as to the presence of a dense molecular cloud towards the north of the nebula. EVLA observation at 1.4 GHz made by \citet{2012BAAA...55..179C} reveals, in fact, a nearby molecular cloud with a density of $\sim$ 400 cm$^{-3}$. At all wavelengths the nebula emission is observed to extend towards the south of the pulsar. The reason could be that in the past the external part of the supernova shell interacted with the nearby molecular cloud, leading to a relatively fast formation of a reverse shock on the northern side of the nebula. The recoil of this reverse shock forced the nebula to expand more towards the southern side.

\subsection{Energy dependent analysis of HESS\,J1825$-$137}
\subsubsection{Energy-dependent extension analysis with a 2D-Gaussian template}
\label{ext_2DGaussian}
During the analysis of the energy-dependent extension, we fixed the spectra of HESS\,J1825$-$137 and of the nearby sources using the resulting model from the initial spectral analysis on the whole energy range (see Sect. \ref{sec:analysis_procedure}).
The normalisation of the isotropic plus diffuse components are instead left as free parameters of the fit. In this part, we performed the localisation and extension analysis separately in each energy band, using 2 bands per decade between 1 and 100\,GeV, and a single band between 100\,GeV and 1\,TeV, keeping the internally 8 bins per decade in the science tools analysis.
Before performing the extension analysis, for each energy band the localisation is again optimised. The extension is then estimated by fitting a 2D-Gaussian template in each energy band and simultaneously refitting the source position. 
We estimated the systematic error of the extension $\sigma_{R,sys}$, due to the choice of the diffuse model, as: 
\begin{equation}
\label{eq:err_ext}
\small
\sigma_{R,sys} = \sqrt{(\mathrm{R}_{D1}-\mathrm{R}_{D2})^2 + (\sigma_{R_{D1}}-\sigma_{R_{D2}})^2 + \Delta^2} \; ,
\end{equation}
\noindent where $\Delta$ is the distance between the source centroids for the two different diffuse models that we considered in our analysis. 
Table \ref{table_extent} contains the results of the extent measurements performed with the 2D-Gaussian template in 2 logarithmic bands per decade between 1 and 100\,GeV and in a single band between 100\,GeV and 1\,TeV . 
\begin{table*}[h]
\small
\centering
\begin{tabular}{c|c|cccc|c}
Energy (GeV) &  Diff. & Ext. ($R$) ($^{\circ}$) & TS$_{ext}$ &  R.A. & Dec. & $\sigma_{Ext,syst}$($^{\circ}$) \\
\hline
1 -- 3 & D1 & 1.11(10) & 132 & $18^h23^m12^s(4^m48^s)$ & $-13^{\circ}30'35''(5'24'')$ & 0.18\\
	 & D2 & 1.00(12) & 125 & $18^h22^m39^s(4^m48^s)$ & $-13^{\circ}24'00''(4'48'')$  &  \\
3 -- 10 & D1 & 1.43(11) & 270 & $18^{h} 23^{m} 43^{s} (4^{m} 12^{s}$) & $-13^{\circ} 41'23''(4'48'')$ & 0.12 \\ 
     & D2 & 1.39(8) & 312 &  $18^h23^m24^s(3^m36^s)$ & $-13^{\circ}43'12''(3'36'')$ & \\ 
10 -- 32 & D1 & 1.47(8) & 357 & $18^h24^m43^s(4^m12^s)$ & $-13^{\circ}55'11''(4'48'')$ & 0.13\\
    & D2 & 1.44(10) & 425  &  $18^h24^m17^s(3^m36^s)$ & $-13^{\circ}52'11''(3'36'')$  &  \\
32 -- 100 & D1 & 1.04(9) & 300 & $18^h25^m31^s(3^m36^s)$ & $-14^{\circ}03'36''(4'12'')$ &   0.11\\
    & D2 & 1.12(10) & 344 &  $18^h25^m17^s(4^m12^s)$ & $-14^{\circ}01'47''(4'12'')$ & \\
100 -- 1000 & D1 & 0.84(8) & 232 & $18^h25^m27^s(4^m48^s)$ & $-14^{\circ}00'00''(4'48'')$ & 0.08\\
    & D2 & 0.91(8) & 205 & $18^h25^m14^s(5^m24^s)$ & $-13^{\circ}58'48''(5'24'')$ & \\
\hline
\end{tabular}
\caption{\small \label{table_extent}
Extension and localisation measurements as a function of energy with statistical and systematic errors. The analysis is performed using two different diffuse models, respectively, `D1' which is the diffuse template used in the 4FGL model \citep{2020ApJS..247...33A} and `D2' which is the diffuse template specifically developed for accurate analysis near the Galactic centre \citep{2017ApJ...840...43A}. The extension is characterised by the 68\% containment radius obtained from a 2D-Gaussian template fit. 
We estimated the systematic error of the extension $\sigma_{Ext,syst}$, due to the diffuse model using Eq. \ref{eq:err_ext}.}
\end{table*}

\begin{figure*}[h]
\begin{minipage}{.33\textwidth}
\includegraphics[width=\columnwidth,width=1.1\columnwidth,  trim=15mm 3mm 13mm 12mm,clip=true]{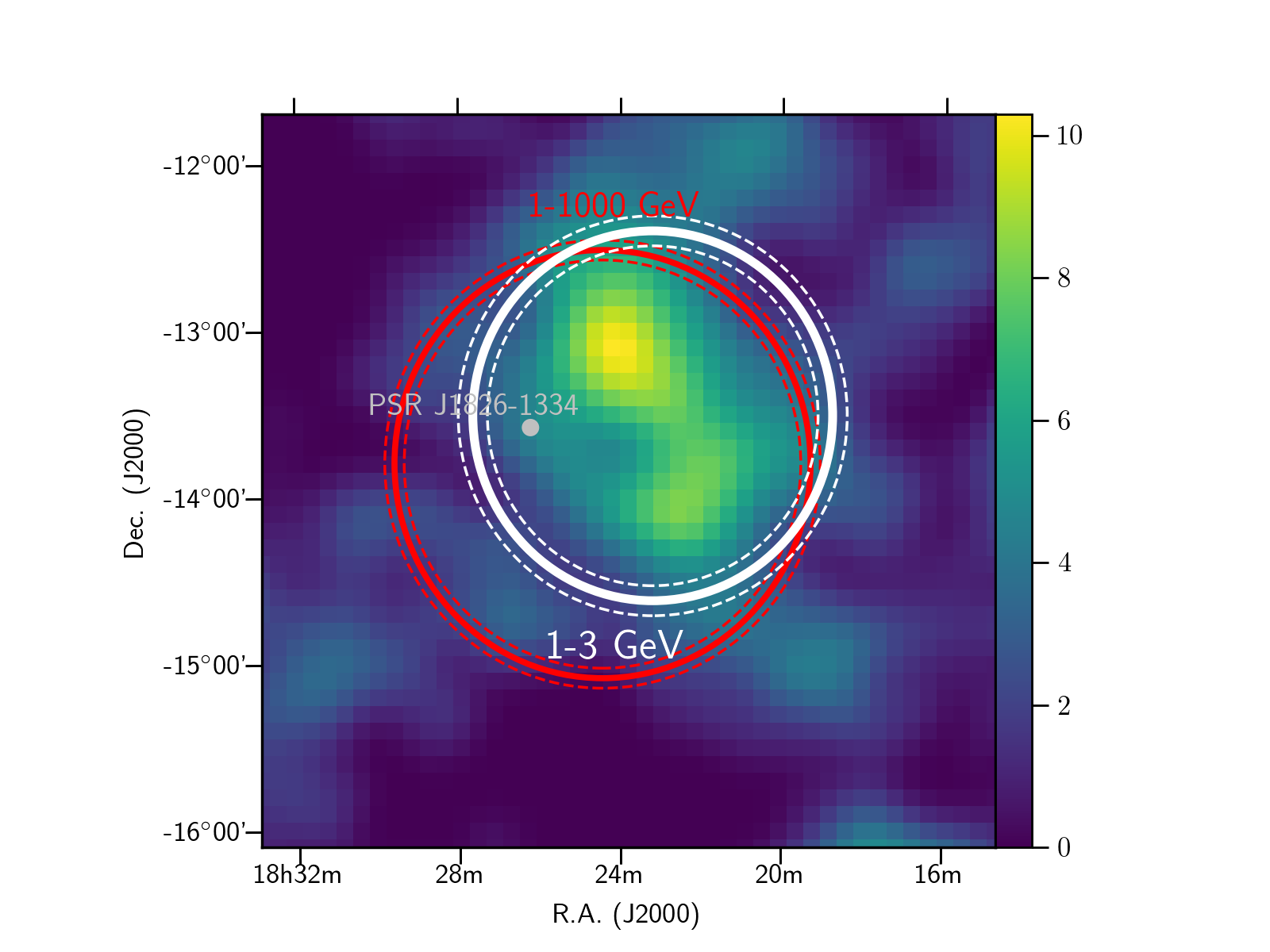}
\end{minipage}
\begin{minipage}{.33\textwidth}
\includegraphics[width=\columnwidth,width=1.11\columnwidth,  trim=15mm 3mm 13mm 12mm,clip=true]{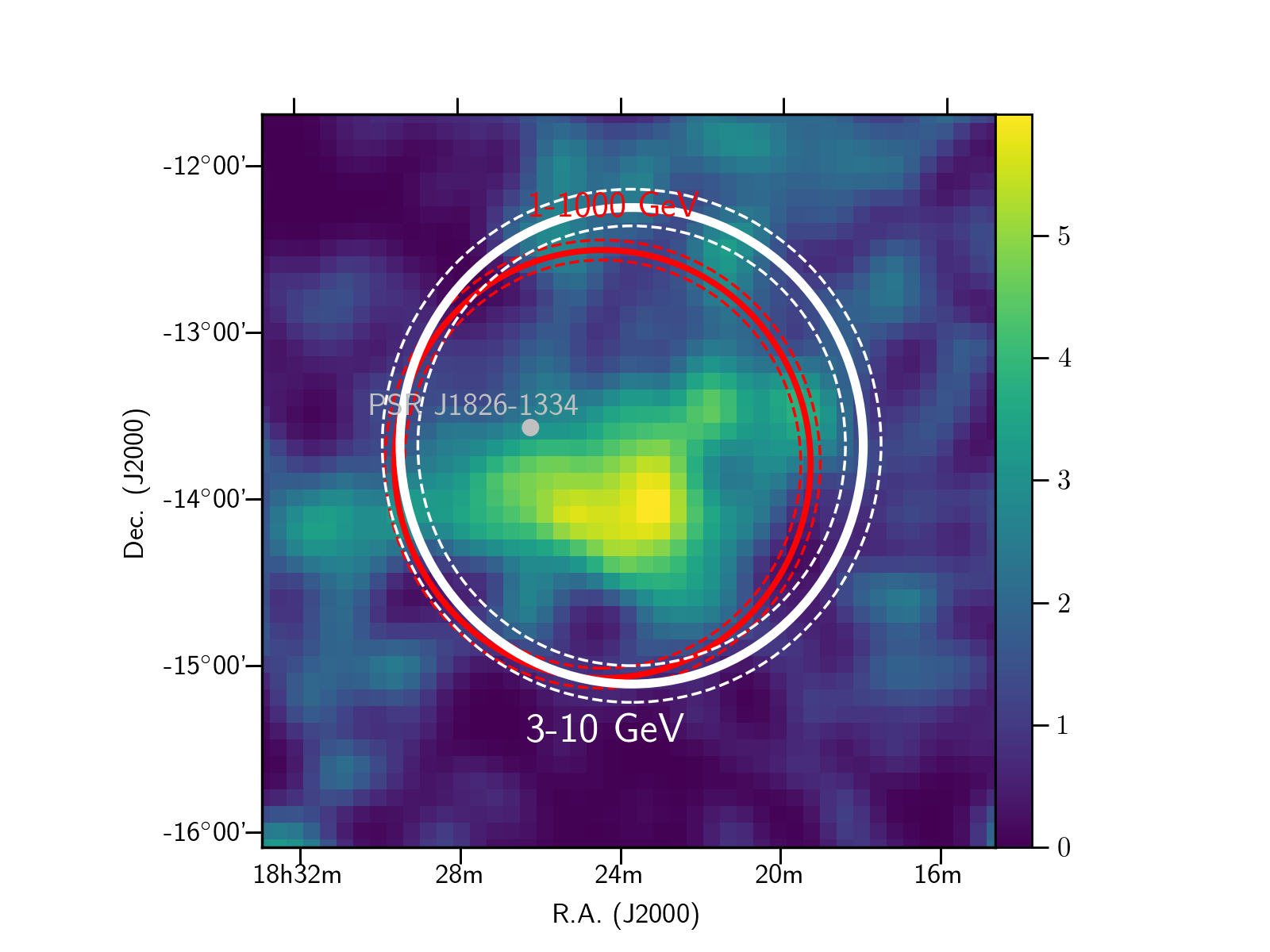}
\end{minipage}%
\begin{minipage}{.33\textwidth}
\includegraphics[width=\columnwidth,width=1.11\columnwidth,  trim=15mm 3mm 13mm 12mm,clip=true]{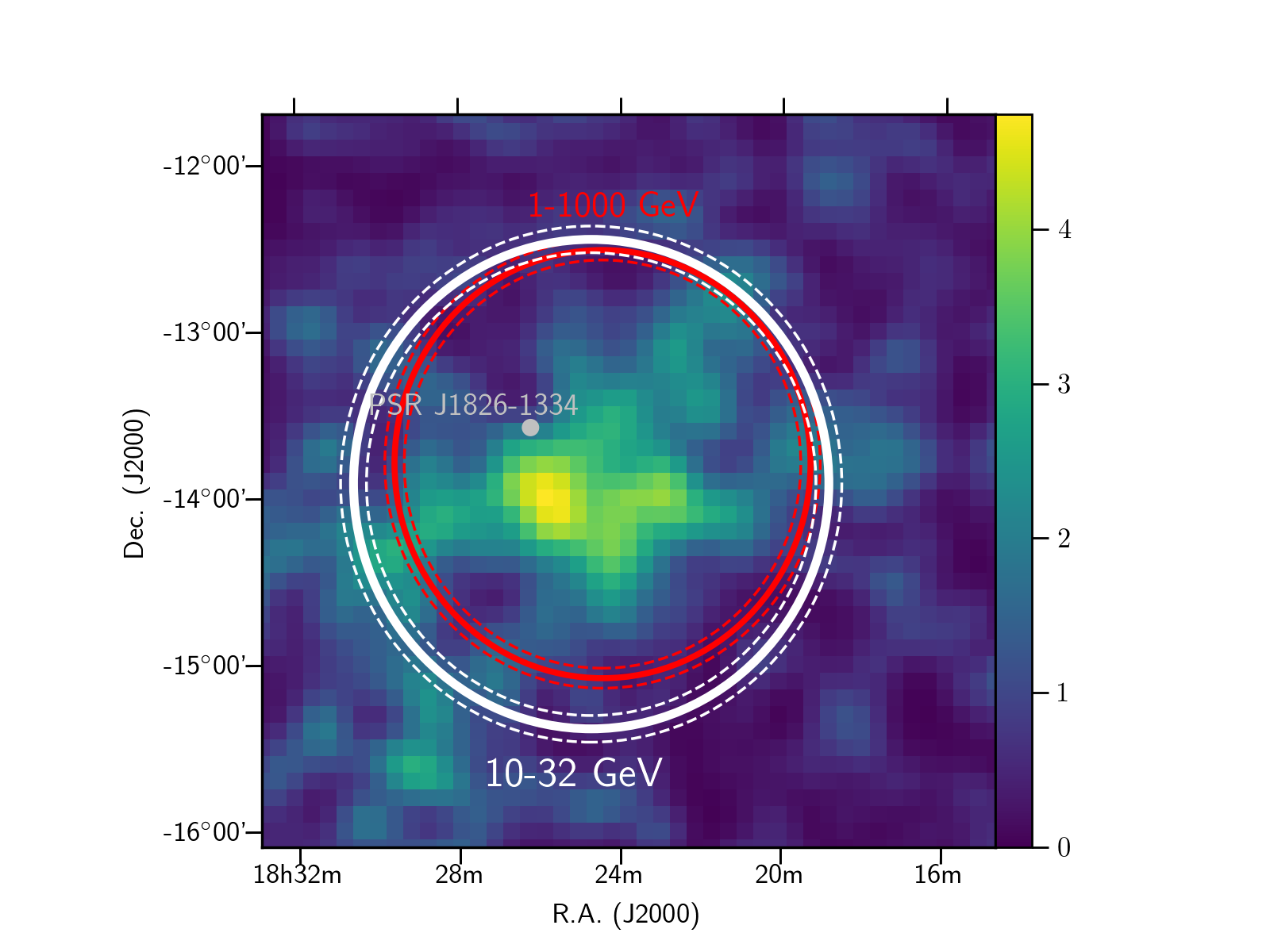}
\end{minipage}%
\\
\begin{minipage}{.33\textwidth}
\includegraphics[width=\columnwidth,width=1.1\columnwidth,  trim=15mm 3mm 13mm 12mm,clip=true]{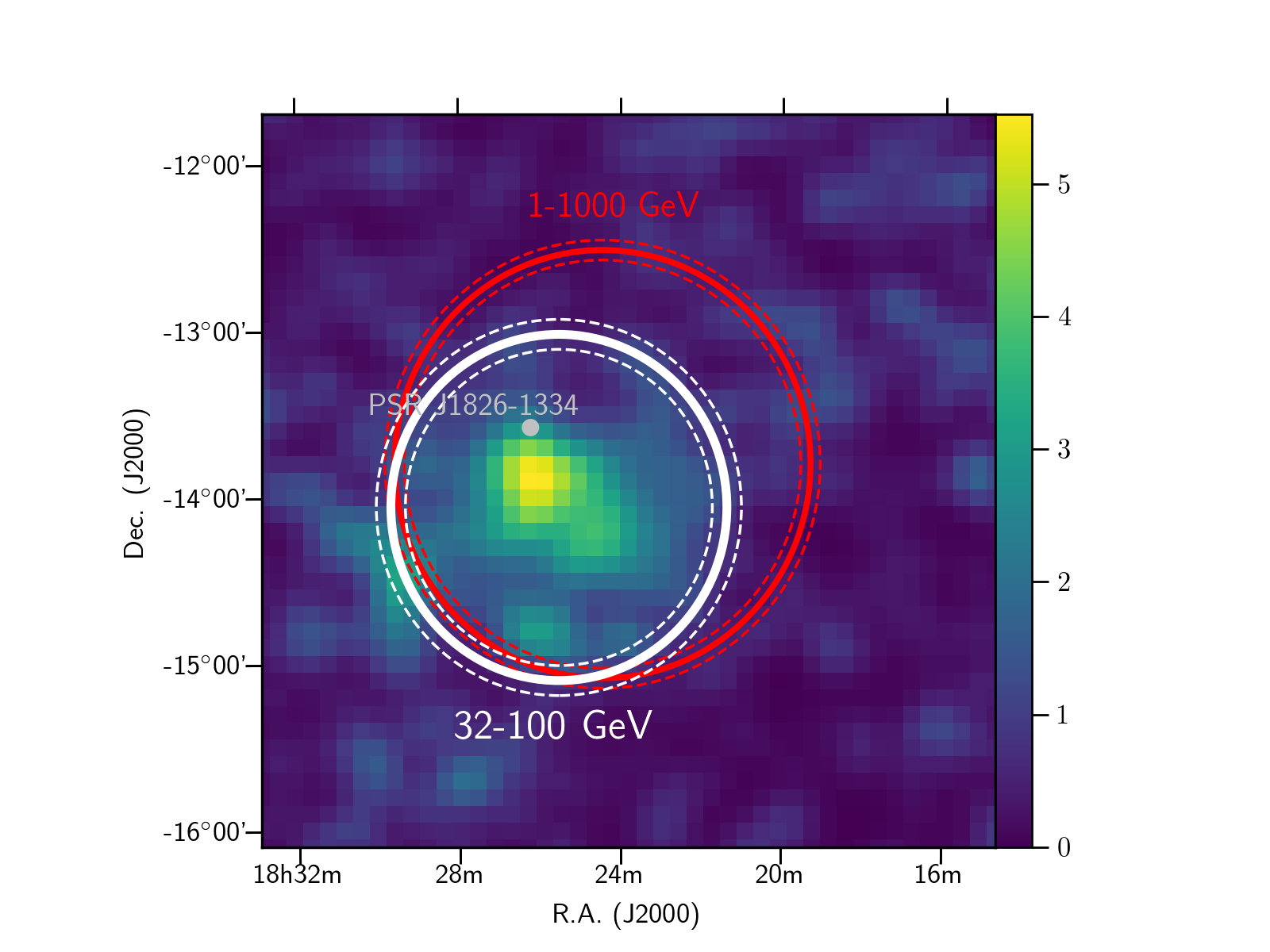}
\end{minipage}%
\begin{minipage}{.33\textwidth}
\includegraphics[width=\columnwidth,width=1.1\columnwidth,  trim=15mm 3mm 13mm 12mm,clip=true]{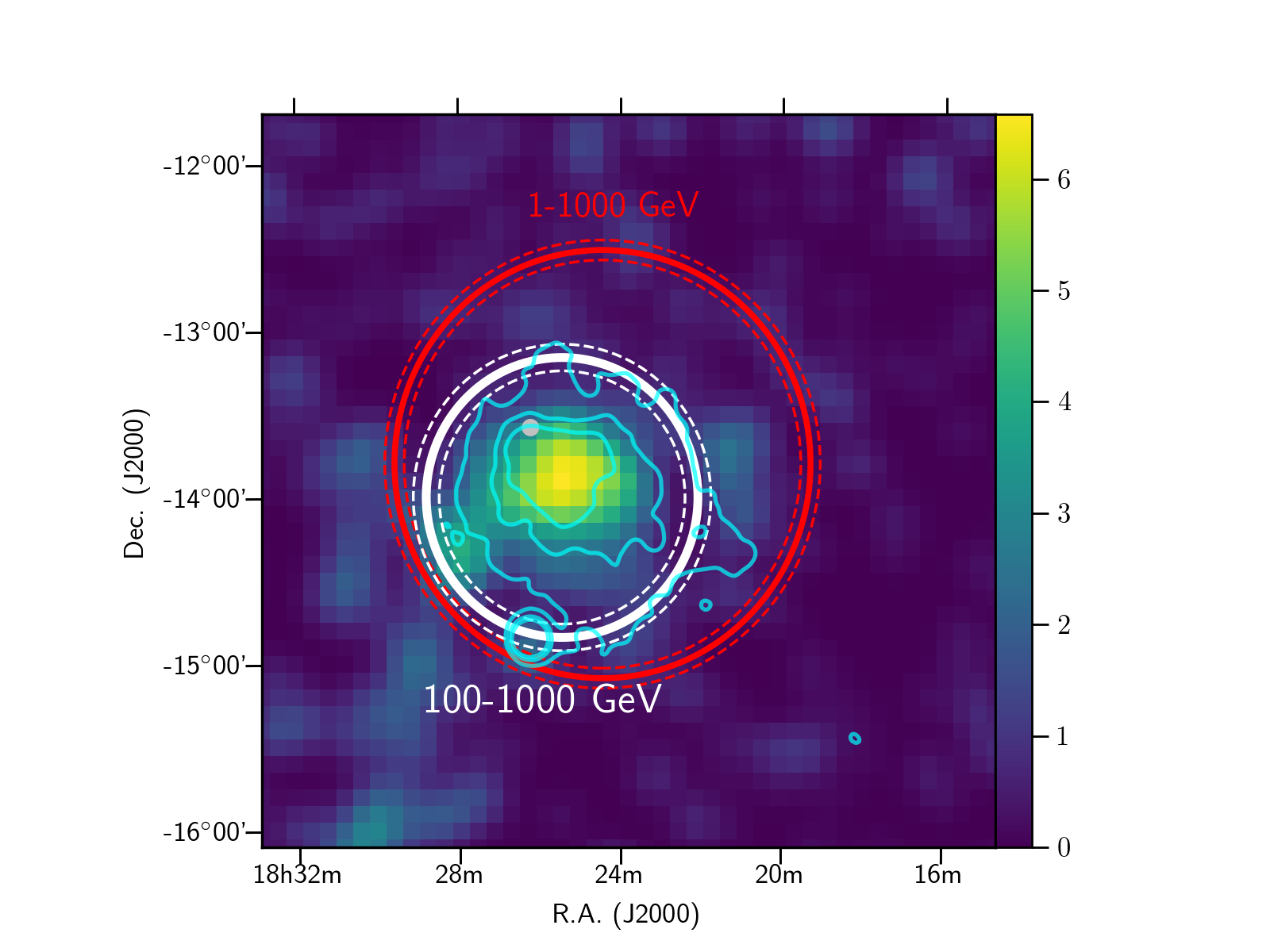}
\end{minipage}%
\caption{\small \label{significance_maps} TS maps (in sigma units), in celestial coordinates, of the region around HESS\,J1825$-$137 for the energy bands: 1--3\,GeV (top left), 3--10\,GeV (top centre), 10--32\,GeV (top right), 32--100\,GeV (bottom left) and 100\,GeV -- 1\,TeV (bottom right).
The TS maps are smoothed with a Gaussian of radius 0.1$^{\circ}$. The white circles represent the extension (solid line) and its statistical uncertainty (dashed lines) determined in the respective energy band.
For comparison, the resulting extension obtained for the entire energy range (1\,GeV -- 1\,TeV) is overlaid with a red line. All extensions shown correspond to the 68\% containment radius. In the 100\,GeV -- 1\,TeV (bottom right) plot, H.E.S.S. significance contours at 5, 10, and 15 $\sigma$, for energies below 1 TeV \citep{2019A&A...621A.116H}, are shown with light-blue lines for comparison. The H.E.S.S. contour also includes the excess of the nearby LS\,5039 source (small circular excess at the southern boarder of the PWN).}
\end{figure*}

The extension estimates ($R$) obtained with the two different diffuse models (D1 and D2), are compatible for each energy band. Similarly the position of the source centroid in the two models are compatible or, in any case, they deviate by a distance which is smaller than the \textit{Fermi}-LAT PSF at the corresponding energy.
The PWN position is observed to vary with energy between the different bands. Moving from the lowest energy band, 1--3\,GeV, to the highest energy one, 100\,GeV -- 1\,TeV, the fitted 2D-Gaussian centroid of the PWN moves towards the current position of the pulsar (see Fig. \ref{significance_maps}).

\subsubsection{Energy-dependent extension analysis with the radial profile method}
\label{radial_profile_analysis}
For comparison, we additionally perform measurements of the nebula extent using the same approach as that adopted in \citet{2019A&A...621A.116H}.
We estimated the extent of the nebula as a function of energy as the radial distance at which the emission in the southern half of the nebula drops to a factor $1/e$ relative to the maximum, starting from the position of the pulsar PSR\,J1826$-$1334 which powers the system.

\begin{figure*}[h]
\centering
\includegraphics[width=9cm,trim=0mm 0mm 0mm 0mm,clip=true]{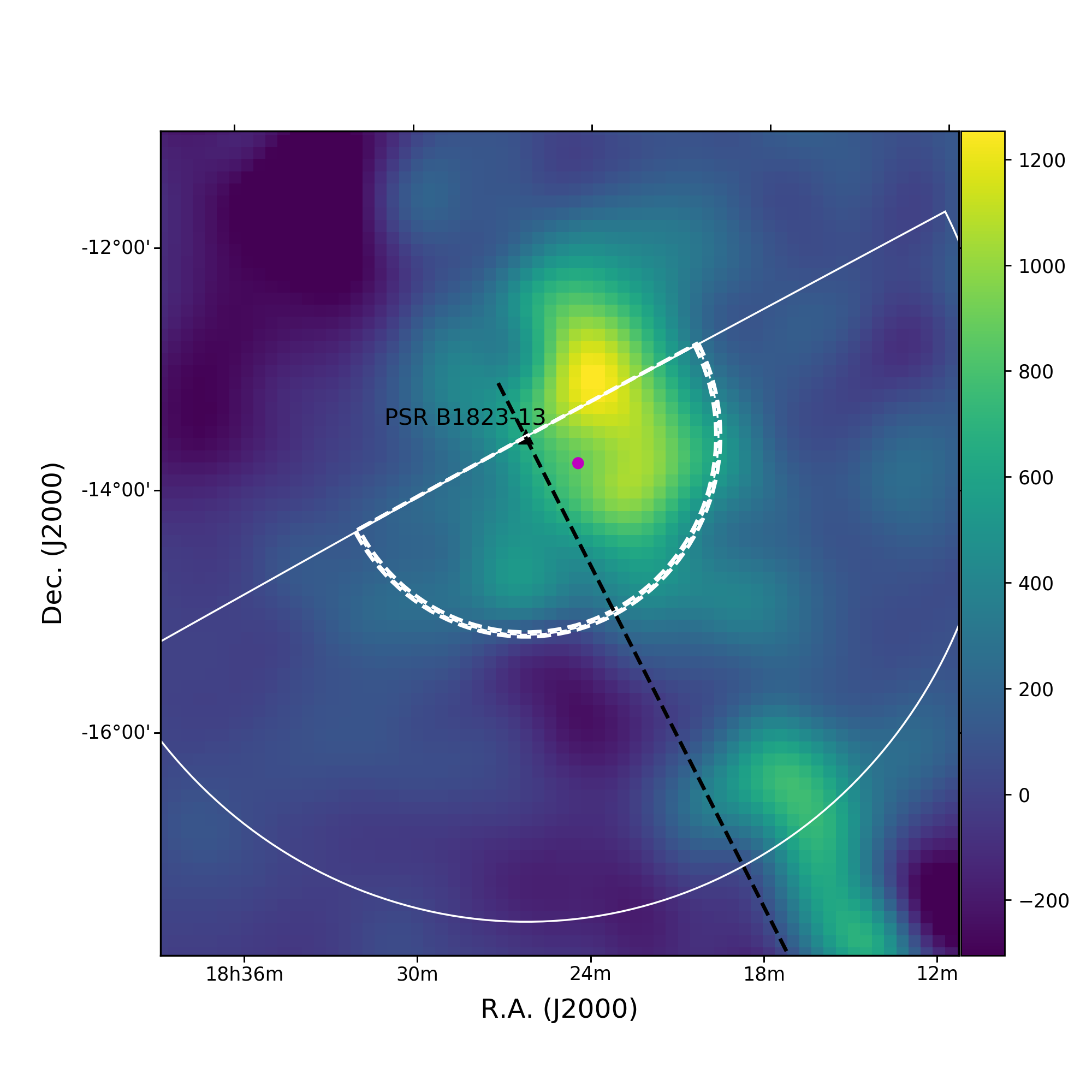}
\includegraphics[width=9cm]{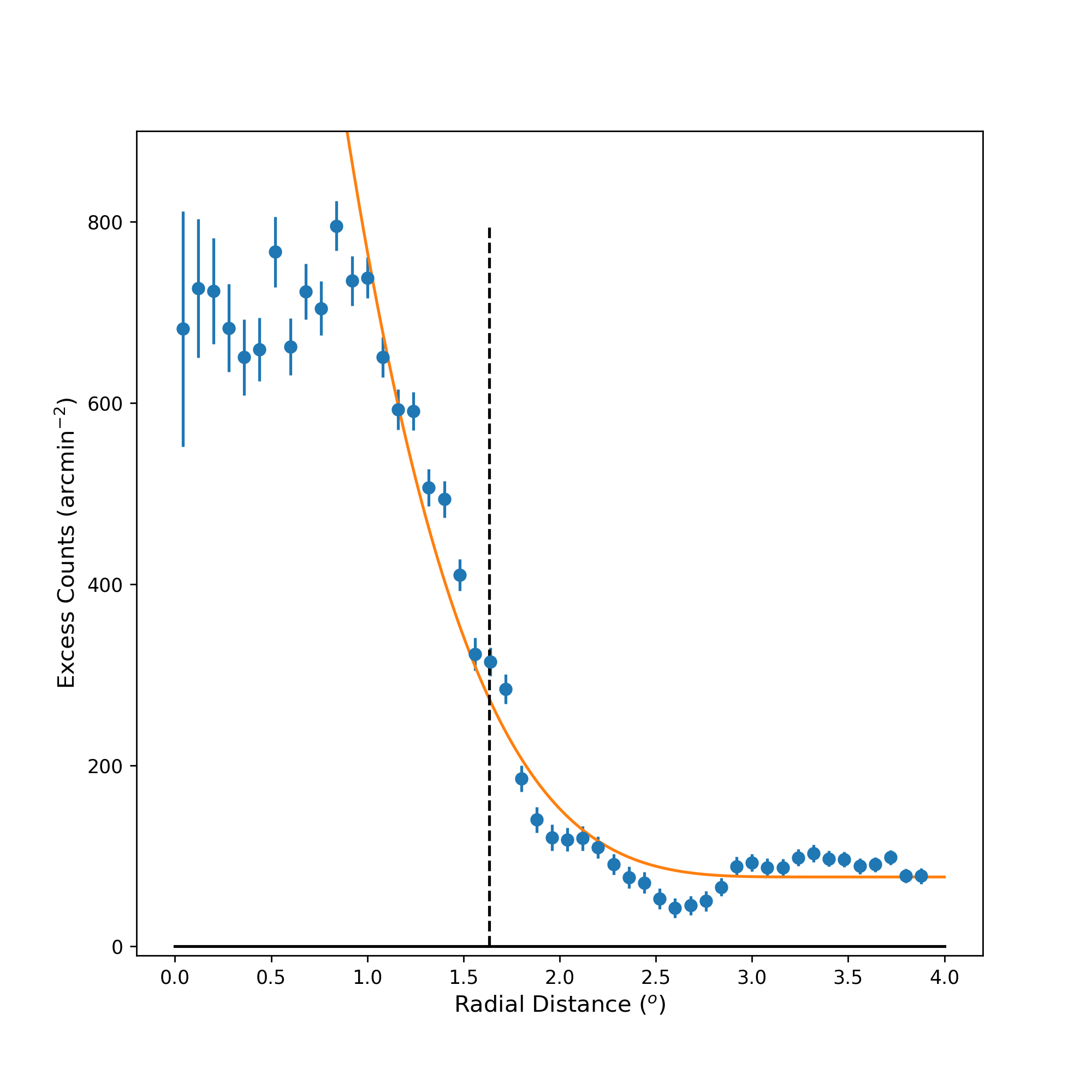}
\caption{Left: TS map (in counts) of the PWN HESS\,J1825$-$137 in the energy range 1\,GeV - 1\,TeV with \textit{Fermi}-LAT, with the region used to extract the radial profile (as used by H.E.S.S.) overlaid in white. The preferred emission direction (`major axis') as found by H.E.S.S., along which the extent is evaluated, is indicated by the black dashed line. The position of the pulsar (black) and best-fit \textit{Fermi}-LAT centre of a 2D Gaussian (magenta) are also indicated. Right: radial profile of the excess counts fit with Eq.\eqref{eq:pol_fit} beyond the peak emission. The characteristic R(1/e) size of the nebula is indicated by a white (black) dashed line in the left (right) hand plot.}
\label{fig:radialprofile}
\end{figure*}
The emission is considered only in one hemisphere due to the strong asymmetry of the source. The orientation of the semi-circular region is the same as that used in the H.E.S.S. analysis, with the major axis of the emission orientated along an angle of $208^\circ$ with respect to the direction of positive declination, as shown in Fig. \ref{fig:radialprofile}.

We estimated the extent of the emission by fitting a polynomial to the radial profile from a minimum radius, out to $4^{\circ}$, using the formula:

\begin{equation} \label{eq:pol_fit}
y(x) = \begin{cases}
a (r-r_{0})^{n} + c , & (x<r_{0}) \\
c, & (x \geq r_{0})
\end{cases} 
\end{equation}

\noindent such that with increasing $r$ the emission decreases out to a distance $r_{0}$  at which it approaches the constant value $c$. The minimum radius of the fit was chosen to be beyond the emission peak, the position of which was determined by a moving average approach. The position of the maximum emission was found to be offset from the pulsar and to vary with energy. The parameter $a$ provides the overall normalisation, whilst the fitted value $r_{0}$ defines the maximum extent (see Table \ref{tab:poly_fit_params} for the values of the fit parameters). 

Since the distance $r_{0}$ was found to be highly sensitive to the order of the polynomial $n$, to mitigate this effect, the extension was taken as the radius at which the fitted function dropped to a fraction ($1/e$) of the peak value, $R_{1/e}$, as a characteristic extent of the nebula. 
This parameter, or indeed any other fixed fraction of the peak value, was found to be stable to the arbitrary choice of the polynomial index $n$, as tested in the range $n=2-5$. We chose a value of $n=4$ in equation \eqref{eq:pol_fit}. 
We evaluated the errors on the extension by performing the fit procedure on 1000 Monte Carlo realisations of the radial profiles in each energy band. 
The radial profiles were generated according to a random number selected from assuming a symmetric Gaussian distribution for each point, with width corresponding to the error on the point.
The nature of the fitted function naturally results in an asymmetric error with larger extensions being more difficult to arise from statistical fluctuations, errors are therefore represented by the 16$^{\rm th}$ and 84$^{\rm th}$ percentiles of the extension distribution.

\subsubsection{Comparison of the 2D-Gaussian and radial profile extent estimates}

The extension of the PWN in the LAT energy range obtained by Fermipy is given as the 68\% containment radius ($R_{68\%}$) of a 2D symmetric Gaussian.
In order to compare the two methods it is possible to approximate the polynomial function used for the radial profile analysis as a simple Gaussian without introducing a large bias.
In this case, for a single Gaussian, the radius at which the function drops to the ratio $1/e$ of the peak value corresponds to $R(1/e)=\sqrt{2} \sigma$.
Consequently, we can compare the extent results obtained with the two different methods (2D-Gaussian and radial profile) using the relation:

\begin{equation}
R(1/e)=\sqrt{2}\frac{R_{68\%}}{\sqrt{-2 \log(1-0.68)}}= 0.937  R_{68\%} \: . 
\label{eq:gausstor1e}
\end{equation}

\noindent Since the extent obtained with the radial profile method is estimated starting from the pulsar position and considering the semi-circular region oriented along the major axis, to directly compare the results of two methods, we found from the 2D-Gaussian results the distance from the pulsar along the major axis at which the $R(1/e)$ of equation \eqref{eq:gausstor1e} intersects the major axis.
This correction accounts for the differences in the 2D-Gauss Ext. between Tables \ref{table_extent} and \ref{table_extent_comp}.

\noindent Table \ref{table_extent_comp} presents the extension results obtained with the two different methods: 2D-Gaussian template and radial profile analysis.

\begin{table}[h]
\caption{\small \label{table_extent_comp}
Extent measurements (using diffusion model D1), in the radial profile $R(1/e)$ format, as a function of the energy for the 2D-Gaussian template (corrected for corresponding extent along the major axis) and the radial profile method using a polynomial fit (Eq.\ref{eq:pol_fit}).}
\small
\centering
\begin{tabular}{ccc}
Energy & 2D-Gauss Ext. & Radial Prof. Ext.\\
 (GeV) & ($^{\circ}$ ) & ($^{\circ}$ )\\
\hline
\hline
1 -- 3 & $1.43 \pm 0.21$ & $ 1.72  ^{+0.18}_{-0.19} $\\ 
3 -- 10 & $1.67 \pm 0.16$ & $ 1.48^{+0.12}_{-0.13}$ \\ 
10 -- 32 & $1.83 \pm 0.15$ & $ 1.33^{+0.13}_{-0.14}$ \\ 
32 -- 100 & $1.17 \pm 0.14$ & $ 1.09^{+0.12}_{-0.16}$\\ %
100 -- 1000 & $0.95 \pm 0.11$ & $ 0.92^{+0.08}_{-0.13}$ \\
1 -- 1000 & $ 1.39 \pm 0.10 $ & $ 1.63  ^{+0.08}_{-0.08} $\\%
\hline
\end{tabular}
\end{table}

\begin{figure}[h]
\centering
\includegraphics[trim=10mm 20mm 10mm 20mm,clip,width=\columnwidth]{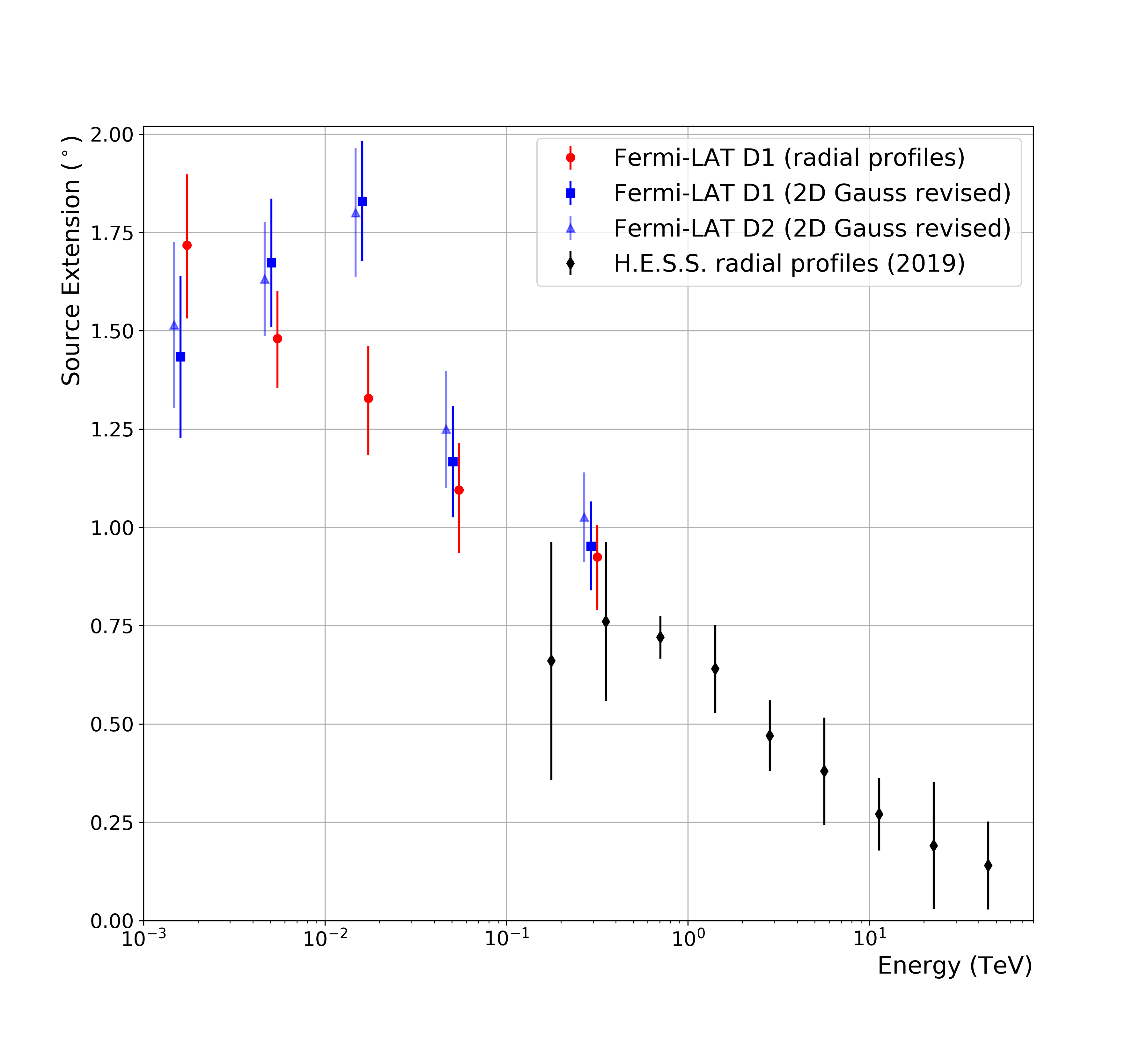}
\caption{\small \label{alt_extension} Nebula extent as a function of energy showing results from \citet{2019A&A...621A.116H} and this analysis.}
\label{fig:exten}
\end{figure}

The results from the 2D-Gaussian, including the offset of the centroid from the pulsar position, are broadly compatible with the results obtained from the radial profile (see Fig. \ref{alt_extension}), except for the energy band between 10 and 30 GeV. The differences may be due to the different regions considered in the extension analysis: the radial profile method takes the asymmetry of the source into consideration and performs the analysis only in the southern hemisphere, whilst the 2D-Gaussian assumes a symmetric source emission. 
Another possible reason is the different treatment of the extension as $R(1/e$); in the case of the radial profiles, the peak value is obtained from the excess counts independent of the fitted function; however, for the 2D-Gaussian, $R_{68\%}$ and the derived $R(1/e)$ relies on the normalisation of the fit. If the normalisation of the Gaussian under- or overestimates the true peak value in the excess counts, which may occur due to the smoother curvature of the Gaussian function, then the distance $R(1/e)$ is shifted accordingly.

Our \textit{Fermi}-LAT results extend the information of the energy-dependent extension of the PWN down to an energy of 1\,GeV.
The \textit{Fermi}-LAT and H.E.S.S. extents for the common energy band (100\,GeV - 1\,TeV) are compatible. The \textit{Fermi}-LAT results reveal a continuous increase of the nebula extent towards lower energies, with a trend similar to that seen by H.E.S.S. above 100\,GeV. 
The possible turnover observed by H.E.S.S. around 300 GeV appears to be ruled out by the LAT results at lower energies, leaving to the possibility of a turnover around few GeV.

\subsection{\textit{Fermi}-LAT spectral energy distribution}
For the analysis of the source spectrum, we divided the photon events into 24 logarithmically distributed energy bins between 1\,GeV and 1\,TeV. We generated the spectral energy distribution (SED), using a 2DGaussian template for the source, by doing a spectral analysis in the various energy bands with their respective spatial parameters (see Table \ref{table_extent} in  Sect. \ref{ext_2DGaussian}).
We corrected for the jumps in the spatial model at the boundaries of the larger energy bands, by slightly enlarging the energy bands and averaging the flux estimates for the overlapped energy bins, as well as by increasing the error bars on the original results to include the uncertainty on the spatial modelling in the interim energy bands.
The diffuse background was were left free to vary.

The SED is fitted with a LogParabola model (see Eq. \ref{eq:logparabola}) as well as with a Broken Power Law (Broken PL) function, which is not a very natural or smooth model, but provides a better estimate of the break energy, $E_b$. The function used for the Broken PL law is:

\begin{equation}
\label{broken_pl}
\dfrac{dN}{dE} = N_{0} \times \begin{cases} 
	(\frac{E}{E_{b}})^{- \Gamma_{1}} \; \mathrm{if} \: E<E_{b}
	\\
	(\frac{E}{E_{b}})^{- \Gamma_{2}} \; \mathrm{otherwise} \:.
	\end{cases}
\end{equation}

\begin{table*}[h]
\small
\centering
\begin{tabular}{ccc|ccc}
LogParabola & & & Broken PL & & \\
\hline
Parameter &  \textit{Fermi} &  \textit{Fermi} $+$ H.E.S.S. & Parameter &  \textit{Fermi} &  \textit{Fermi} $+$ H.E.S.S. \\
\hline
\hline
$\alpha$ &  1.96 $\pm$ 0.68 & 2.15 $\pm$ 0.05 & $\Gamma_{1}$ & 1.70 $\pm$ 0.04 & 1.69  $\pm$ 0.03\\
$\beta$ &  0.046  $\pm$ 0.013 & 0.075 $\pm$ 0.002 & $\Gamma_{2}$ &  2.29 $\pm$ 0.15 & 2.51  $\pm$ 0.01\\
$E_{0}$ (GeV) &  145 $\pm$ 54 & 154 $\pm$ 38 & $E_{b}$ (GeV) & 115 $\pm$ 8 & 114 $\pm$ 13 \\
$N_{0}$  & 6.67 $\pm$ 0.93 & 5.02 $\pm$ 0.20 & $N_{0}$  & 8.28 $\pm$  0.64 & 8.37 $\pm$  0.47\\
$\chi^{2}$/ndf & 13/20 & 40/34 & $\chi^{2}$/ndf & 10/20 & 71/34\\
\hline
\end{tabular}
\caption{\small \label{table_sed}
Best fit parameters for the SED of HESS\,J1825$-$137 (see Fig. \ref{spectra_hess_j1825}). with a LogParabola and Broken PL models. The parameter normalisation $N_{0}$ is in units of ($10^{-11}$ erg cm$^{-2}$ s$^{-1}$). The `\textit{Fermi}' and `\textit{Fermi} $+$ H.E.S.S.' columns contain the fit results obtained with only the \textit{Fermi}-LAT data and with both \textit{Fermi}-LAT and H.E.S.S. data points respectively.}
\end{table*}

\begin{figure}[h]
\centering
\includegraphics[width=1.1\columnwidth]{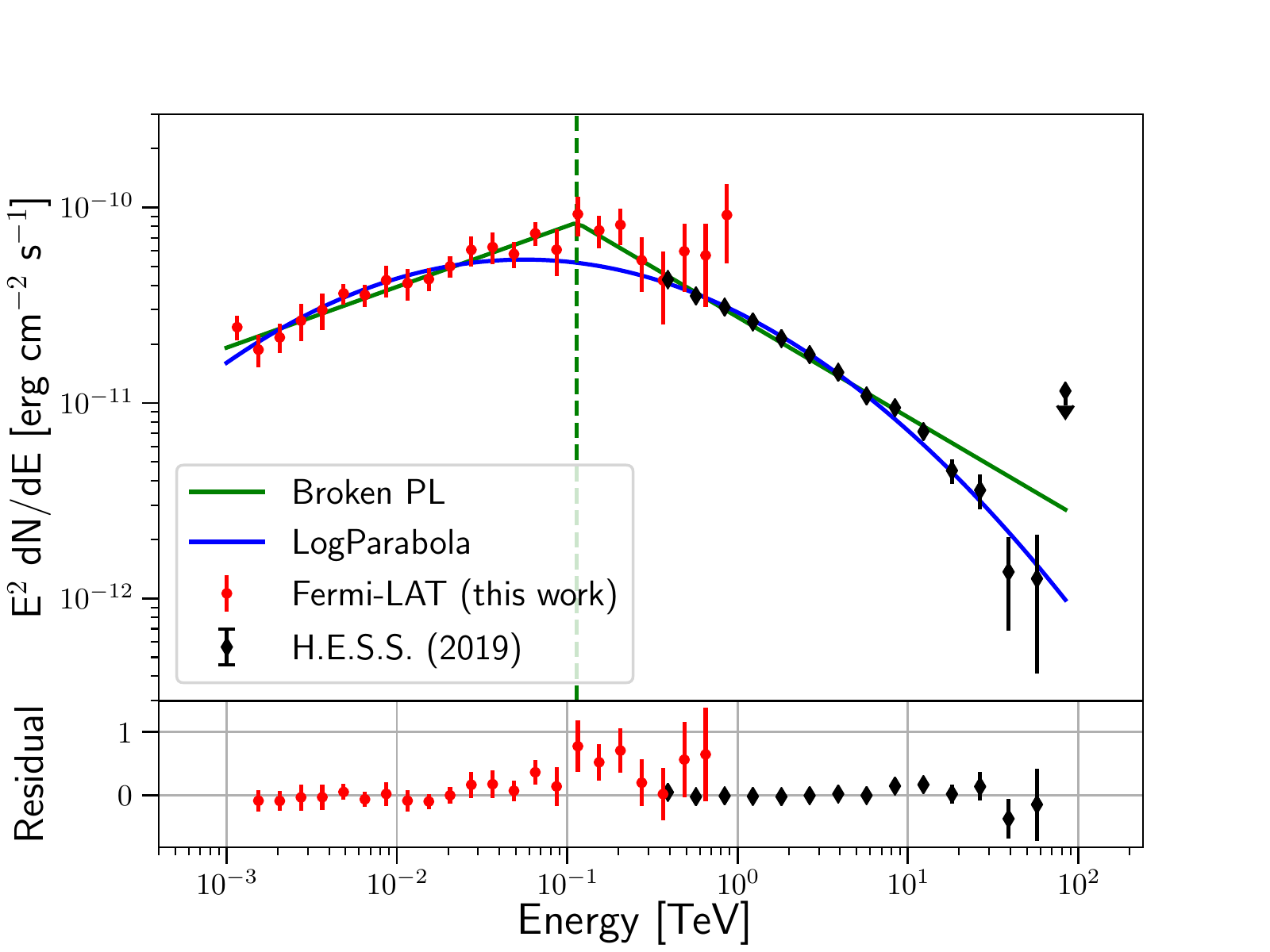}
\caption{\small \label{spectra_hess_j1825}
Combined spectra of the PWN HESS J1825$-$137 with the spectral measurements obtained in this work (red points) using 11.6 years of \textit{Fermi}-LAT data from 1\,GeV to 1\,TeV and the H.E.S.S. results for the 100\,GeV -- 90\,TeV energy range (black points). The \textit{Fermi}-LAT flux points were obtained doing a spectral analysis in the various energy bands with their relative spatial model. The combined SED has been fitted with both a LogParabola (blue line) and a Broken PL (green line). The vertical line corresponds to the energy break $E_{b}$ of the Broken PL model. The bottom panel shows the normalised residual between the data and the LogParabola model.}
\end{figure}

\noindent The fit results for the LogParabola and Broken PL models are reported in the `\textit{Fermi}' column of Table \ref{table_sed}. The `\textit{Fermi} + H.E.S.S.' column of the Table \ref{table_sed} contains the resulting spectral information for the PWN HESS J1825$-$137 from a joint fit to the energy flux points derived independently from \textit{Fermi}-LAT and H.E.S.S. data \citep{2019A&A...621A.116H}. The LogParabola is found to be preferred ($\chi^{2}/\mathrm{ndf} \sim 1 $, with $\mathrm{ndf}=35$) for both the \textit{Fermi} only and the combined spectral fits, and it nicely describes the spectra (see Fig. \ref{spectra_hess_j1825} for the combined SED). The Broken PL model, which fails to describe completely ($\chi^{2}/\mathrm{ndf} > 2$) the spectral results above 10\,TeV due to sharp cutoff, returns a energy break estimate of about 115\,GeV.
  
\section{Modelling of the Nebula}
We modelled the combined GeV-TeV spectral energy distribution of the nebula as Inverse Compton (IC) scattering from a leptonic particle population. Several packages exist for spectral modelling with complementary features. We use both the NAIMA package \citep{naima} for a single zone model due to the statistical fitting methods available, and the modular GAMERA package \citep{Hahn:2015hhw} for a multi-zone model due to the enhanced flexibility offered. 
The two models were found to be consistent except for an offset in the flux and magnetic field constraint, which arise from the different ambient radiation fields used. Whereas a lookup table for the total radiation fields from the model of \citet{2017MNRAS.470.2539P} could be used directly in GAMERA, a black-body approximation to this model was used with the NAIMA package.

\subsection{SED modelling with NAIMA}
\begin{figure}[h]
\centering
\hspace*{-0.4cm}
\includegraphics[width=\columnwidth]{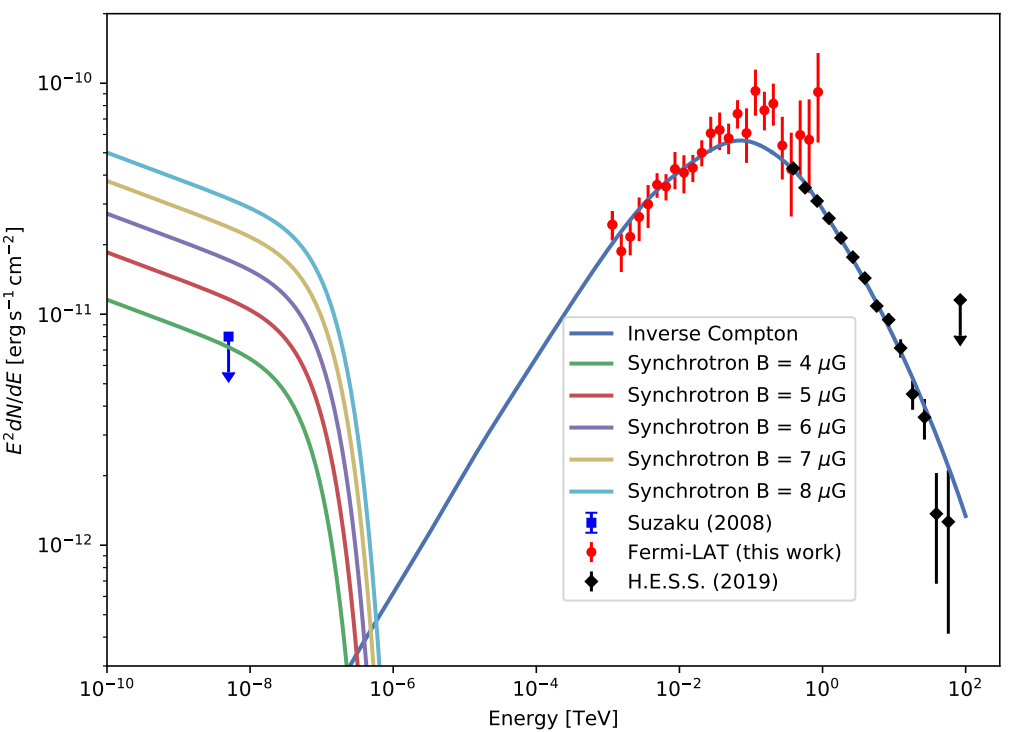}
\caption{\small \label{Naima_model} Results of the fitted IC NAIMA SED to the \textit{Fermi}-LAT data and H.E.S.S. data, compared with the measured \textit{Fermi}-LAT (red circle) and H.E.S.S. (black diamond) data points. A synchrotron component is computed from the electron distribution supposing 4, 5, 6, 7, and 8 $\mu$G, and compared with \textit{Suzaku} data (blue square). The electron distribution parameters are presented in Table \ref{tab:Naima_fit_params}.}
\end{figure}

To test if a simple analytic electron distribution can explain the combined spectra of H.E.S.S. and \textit{Fermi}-LAT, we used the NAIMA python package.  We considered a leptonic population of particles, from an energy of $1$ GeV up to $510$ TeV, producing $\gamma$-ray by IC scattering. The radiation fields at the galactic position of HESS J1825$-$137 are obtained using the parametrization of \citet{2017MNRAS.470.2539P}. They are modelled as five black body components, with different temperatures and energy densities $\varepsilon$ that correspond to Cosmic Microwave Background (CMB, T$=2.72$ K, $\varepsilon=0.26$ eV cm$^{-3}$), Far Infra-Red (IFR, T$=43.07$ K, $\varepsilon=0.78$ eV cm$^{-3}$), Infra-Red (IF, T$=238.4$ K, $\varepsilon=0.17$ eV cm$^{-3}$), VISible (VIS, T$=3493$ K, $\varepsilon=1.8$ eV cm$^{-3}$), and Ultra-Violet (UV, T$=19840$ K, $\varepsilon=0.17$ eV cm$^{-3}$). 
The dominant contributors to the IC $\gamma$-ray flux are the FIR for energy between $3$ GeV and $15$ TeV, and the CMB otherwise. The contribution of the UV is negligible compared to the other radiation fields. The particle population is assumed to follow a broken power-law, its parameters are fitted using the MCMC method implemented in NAIMA. The normalisation of the broken power-law is derived from the total energy of the electrons $W_e$ assuming that the source is located at a distance of $4$\,kpc. 
The results are presented in Fig. \ref{Naima_model} and the parameters of the model are given in Table \ref{tab:Naima_fit_params}. This single population model is able to explain the complete range of energy covered by H.E.S.S. and \textit{Fermi}-LAT data. We used this distribution in order to model the synchrotron component assuming different values of the magnetic field which we compare with X-ray data obtained by \textit{Suzaku} \citep{suzaku}. The X-ray flux point corresponds to the sum of the X-ray emission from a comparatively small region ($\lesssim 15'$) around the pulsar,
re-scaled for a comparable opening angle to that of the $\gamma$-ray analyses, with the X-ray emission assumed negligible outside of this region.
This can be treated as an upper limit on the total X-ray flux averaged over the much larger region probed by the $\gamma$-ray emission. We found that the maximum magnetic field allowed by X-ray observations using this model is $\sim 4 \mu$G.

\begin{table}[h]
\caption{\small \label{tab:Naima_fit_params}
 Derived parameters of the broken power law model. $\Gamma_1$ and $\Gamma_2$ are the first and second power law index, $E_b$ is the break energy and $W_e$ is the total energy of the electron population.}
\small
\centering
\begin{tabular}{cc}
Parameter & H.E.S.S. and \textit{Fermi}-LAT \\
\hline
\hline
$W_e$ ($10^{49}$ erg) & $2.33^{+1.00}_{-0.64}$\\
$\Gamma_1$ & $2.02^{+0.15}_{-0.19}$ \\
$\Gamma_2$ & $3.23^{+0.02}_{-0.02}$ \\
$E_b$ (TeV) & $0.80^{+0.18}_{-0.14}$ \\
$\chi^2/$ndf & $20.8/34$ \\
\hline
\end{tabular}

\end{table}

\subsection{Multi-zone modelling with GAMERA}
\label{sec:gamera}
\subsubsection{SED modelling}
To simultaneously describe the SED (Fig. \ref{spectra_hess_j1825}) and the variation in extent with energy (Fig. \ref{fig:exten}), we attempted a multi-zone modelling approach using the GAMERA package \citep{Hahn:2015hhw}. A leptonic particle population producing $\gamma$-rays by IC scattering was again assumed, with the radiation fields at the location of the PWN obtained from \citet{2017MNRAS.470.2539P} as for the NAIMA modelling; however, the black-body approximation is not necessary with the GAMERA package enabling a more precise parameterisation to be used. The difference this introduces is small, yet allows slightly higher magnetic field strength values of $5-6\mu$G.

\begin{figure}
    \centering
    \includegraphics[width=\columnwidth]{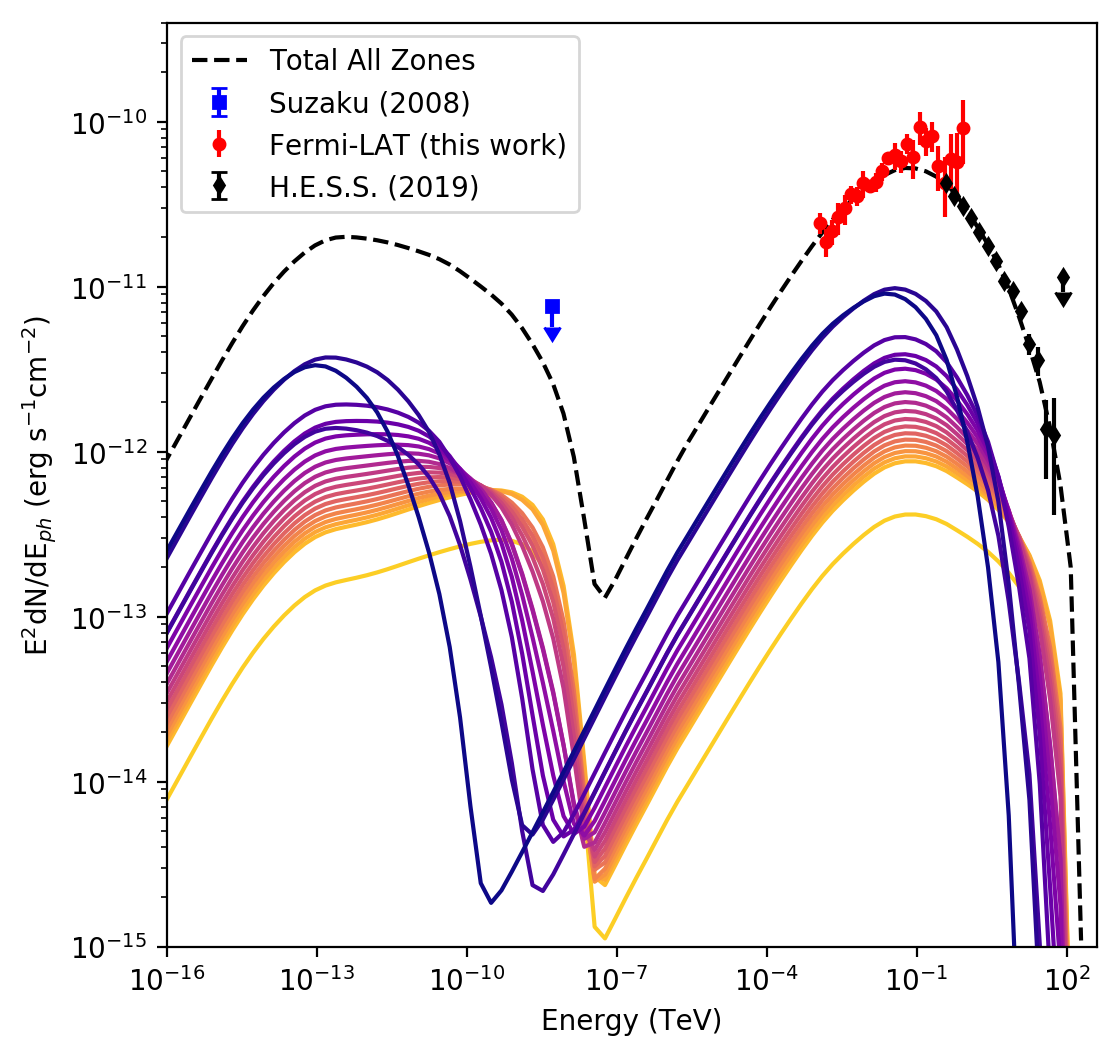}
    \caption{Total SED of the nebula, combining X-ray and $\gamma$-ray data, is described by the summation of 20 zones of particles of different ages using the GAMERA modelling package \cite{Hahn:2015hhw}. The total SED of each zone is shown by a coloured line, from yellow for the youngest and smallest zone, through to blue for the oldest and largest. Model parameters are given in Table \ref{tab:gameramodel}.}
    \label{fig:gamerazones}
\end{figure}

We used a series of particle zones, added with burst-like injection at different times and left these free to evolve in time until the system age is reached. For describing the total $\gamma$-ray emission from the nebula at the current time we used a summation of 20 zones of particles of different ages, evenly split in time.
The model parameters used for the curves shown in Fig. \ref{fig:gamerazones} are given in Table \ref{tab:gameramodel}. Several parameters were constrained to match current values at the present day, with the pulsar characteristic age assumed to be the age of the nebula system. The evolution of the pulsar spin-down luminosity $L$ with time $t$ is described by:

\begin{equation}
    L(t) = (1-\eta)L_0\left(1+\frac{t}{\tau_0}\right)^{-\frac{n+1}{n-1}}~,
\end{equation}

\noindent where $\eta$ accounts for the conversion efficiency, $n$ is the braking index for which a value of 3 was assumed, corresponding to pure magnetic dipole radiation, and $\tau_0$ is the initial spin-down timescale of the pulsar. The spin-period $P$ of the pulsar evolves as:

\begin{equation}
    P = P_0 \left(1+\frac{t}{\tau_0}\right)^{\frac{1}{n-1}}~,
\end{equation}

\noindent from which $\tau_0$ can be determined for an assumed initial spin-down period $P_0$ (a free parameter of the model) using the constrained parameters at the present day age of the system. For the electron population we used a broken power law injection spectrum. 

\begin{table}
\caption{Parameters of the model used to describe the data with the GAMERA modelling package \cite{Hahn:2015hhw}. The constrained parameters are those fixed to measured or derived values at the current time.}
\label{tab:gameramodel}
\small
\centering
\begin{tabular}{c|c|c}
Parameter &  Value & Constrained \\
\hline
\hline
    $T_c$ & 21 kyr & Y \\
    $L(T_c)$ & 2.8$\times10^{36}$ erg/s & Y\\
    $d$ & 4 kpc & Y \\
    $\Gamma_1$ & 1.9 & N\\
    $\Gamma_2$ & 2.8 & N\\
    $E_b$  & 0.3 TeV& N\\
    $E_{\mathrm{max}}$ & 250 TeV & N \\
    $P_0$  & 15 ms & N\\
    $P(TT)$  & 101 ms& Y\\
    $\eta$ & 0.55 & N\\
    $B(T)$ & 5 $\mu$G& Y\\ 
\end{tabular}
\end{table}

The average magnetic field of the nebula was assumed to be $5\,\mu$G at the present day, evolving in time as: 

\begin{equation}
B(t) \propto \left(1+\frac{t}{\tau_0}\right)^{-1}~.    
\end{equation}

As the average magnetic field found for the whole nebula from the models is at $\sim 4-5 \mu$G close to the average ISM magnetic field strength of $\sim 3\mu$G, any spatial dependence is likely to be weak.
In reality, however, the magnetic field is expected to exhibit both temporal and spatial dependence, reducing with increasing distance from the pulsar within the nebula, 
with the strongest evolution in the region nearest the pulsar. Some spatial dependence of the B-field is included in this multi-zone model as a consequence of the different ages and sizes of the emission zones. 
The resulting SED is shown in Fig. \ref{fig:gamerazones}, which is consistent with the available data.

\subsubsection{Radial extent modelling}

In a second step we described the radial extent as a function of energy. We stored the spectra for each zone of the model separately such that the zones could be arbitrarily arranged for the spatial modelling. The zones were treated as expanding shells in space, initially spherically symmetric, with the particle spectra filling the shell volume. 
That is, at a given radial distance $r$ from the pulsar, the line-of-sight depth $d_z$ through a given zone $z$ of radial size $R_z$ is given by:

\begin{equation}
    d_z = \begin{cases}
2\sqrt{R_z^2 - r^2} & (r<R_z) \\
0 & \mathrm{\small{otherwise,}}
\end{cases} 
\end{equation}

\noindent such that the depth through the zone along the line-of-sight decreases towards the edge of the zone and is zero at $r > R_z$. 
The contribution from the spectrum of a given zone to the total emission at a distance $r$ from the pulsar was therefore weighted by $d_z$.

We used the projection of the emission along the line of sight to form an emission profile from the model similar to that of Fig. \ref{fig:radialprofile}. 
This projection was made in multiple energy bands, summing the relevant parts of the zone spectra. 
For determining the radial extent in each energy band we applied the aforementioned procedure (see section \ref{radial_profile_analysis}) of fitting the radial profile to evaluate the distance at which the emission drops to a fraction $1/e$ of the peak value from the model.

\begin{figure}
    \centering
    \includegraphics[width=\columnwidth]{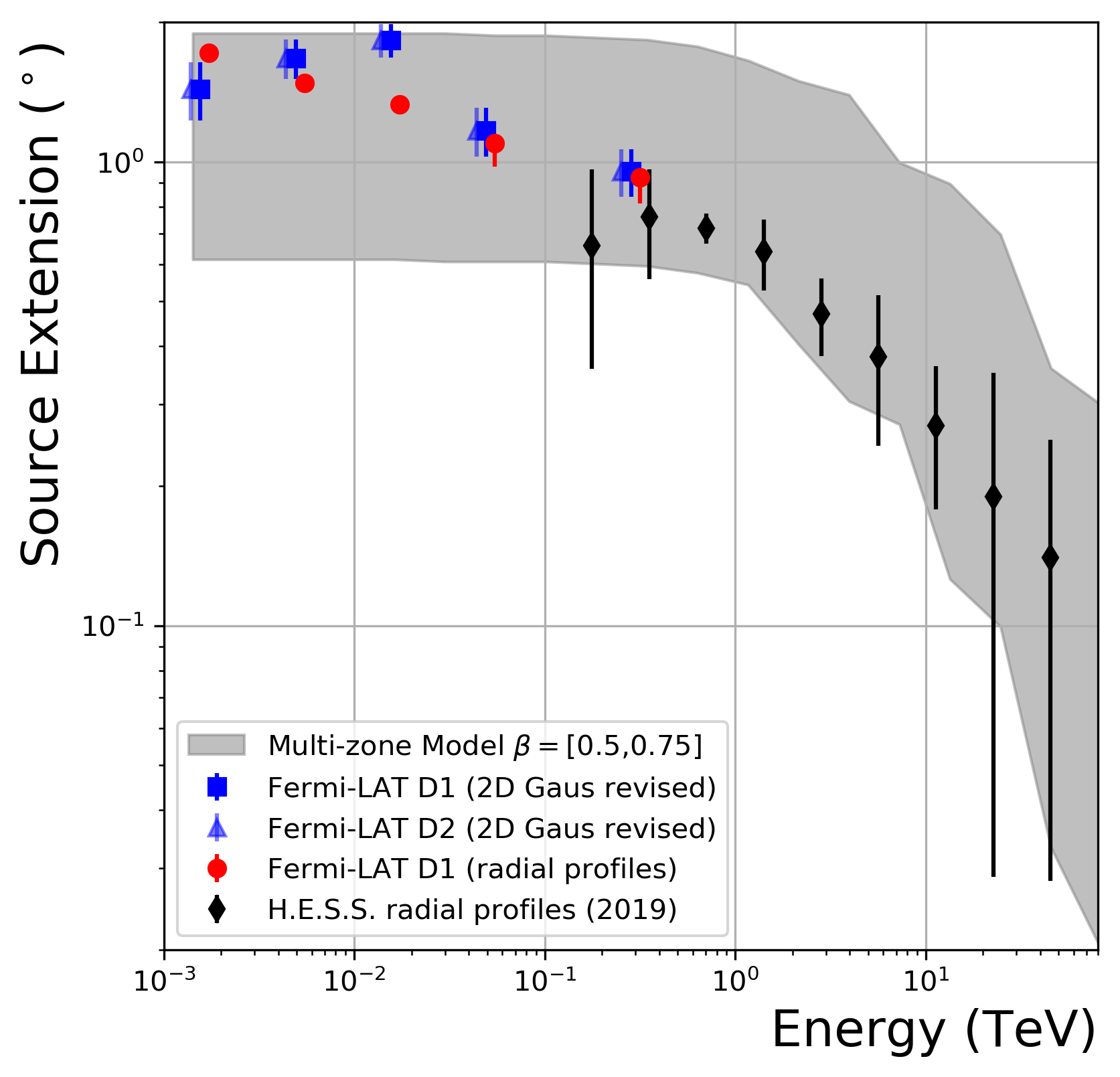}
    \caption{Radial extent of the nebula as a function of energy; the results of the model using GAMERA is indicated by grey shaded region for a compatible range of $\beta$, the index of the radially and time dependency of the velocity profile, between 0.5 (upper edge) and 0.75 (lower edge).}
    \label{fig:gameraRE}
\end{figure}

The energy-dependence of the radial extent could be described using a radially dependent velocity profile $v(r,t)$ with index $\beta$ in the range [0.5,0.75] and initial outflow velocity $v_0=0.03c$ in an advection dominated scenario: 

\begin{equation}
    \centering
    v(r,t) = v_0\left(\frac{r}{r_{\mathrm{max}}}\right)^\beta\left(\frac{t}{T}\right)^{-\beta}~,
    \label{eq:vprof}
\end{equation}
where $r_{\mathrm{max}}$ is the maximum extent of the nebula, for which a value of 150\,pc was assumed, corresponding to a maximum angular extent of $\sim 2^\circ$ based on the extent measurements presented here. The spatial evolution of each of the zones was calculated from a minimum radius of $0.01$pc, corresponding to typical termination shock radii. 
The consistency of this range of $\beta$ values with the experimentally measured extents is shown in Fig. \ref{fig:gameraRE}.

The energy dependent extent of the nebula obtained from this multi-zone model is seen to be approximately constant below a $\gamma$-ray energy of $\sim$0.1\,TeV. 
Assuming that the energy dependent morphology of the nebula is due to electron ageing and cooling, then for un-cooled electrons, the nebula size remains constant in a spherically symmetric scenario.

Whilst a range of values of $\beta$ compatible with previous models of the nebula are consistent with our results, a single $\beta$ value for the full nebula seems not to be able to describe the data, implying either a changing velocity profile over time or non-spherical symmetry. 

\section{Discussion}
We analysed 11.6 years of \textit{Fermi}-LAT data in the energy range between 1\,GeV and 1\,TeV, and we performed, for the first time with \Fermi-LAT data, an energy dependent analysis of the morphology of the PWN HESS J1825$-$137 in the GeV range.

\subsection{System evolution}
\label{system_evolution}
The nebula HESS J1825$-$137 presents a strong energy dependent morphology. The spectral index was seen by \citet{2006A&A...460..365A}, at TeV energies, to soften with increasing distance from the PSR; this implies that the population of electrons in the nebula had travelled and cooled out to large distances. 
The fact that the low-energy electrons at large distances from the pulsar produce the softest spectrum was interpreted to mean that these are the oldest electrons in the system. 

\begin{figure}
    
    \centering 
    \includegraphics[trim=1.0cm 0cm 1.9cm 1cm,clip,width=\columnwidth]{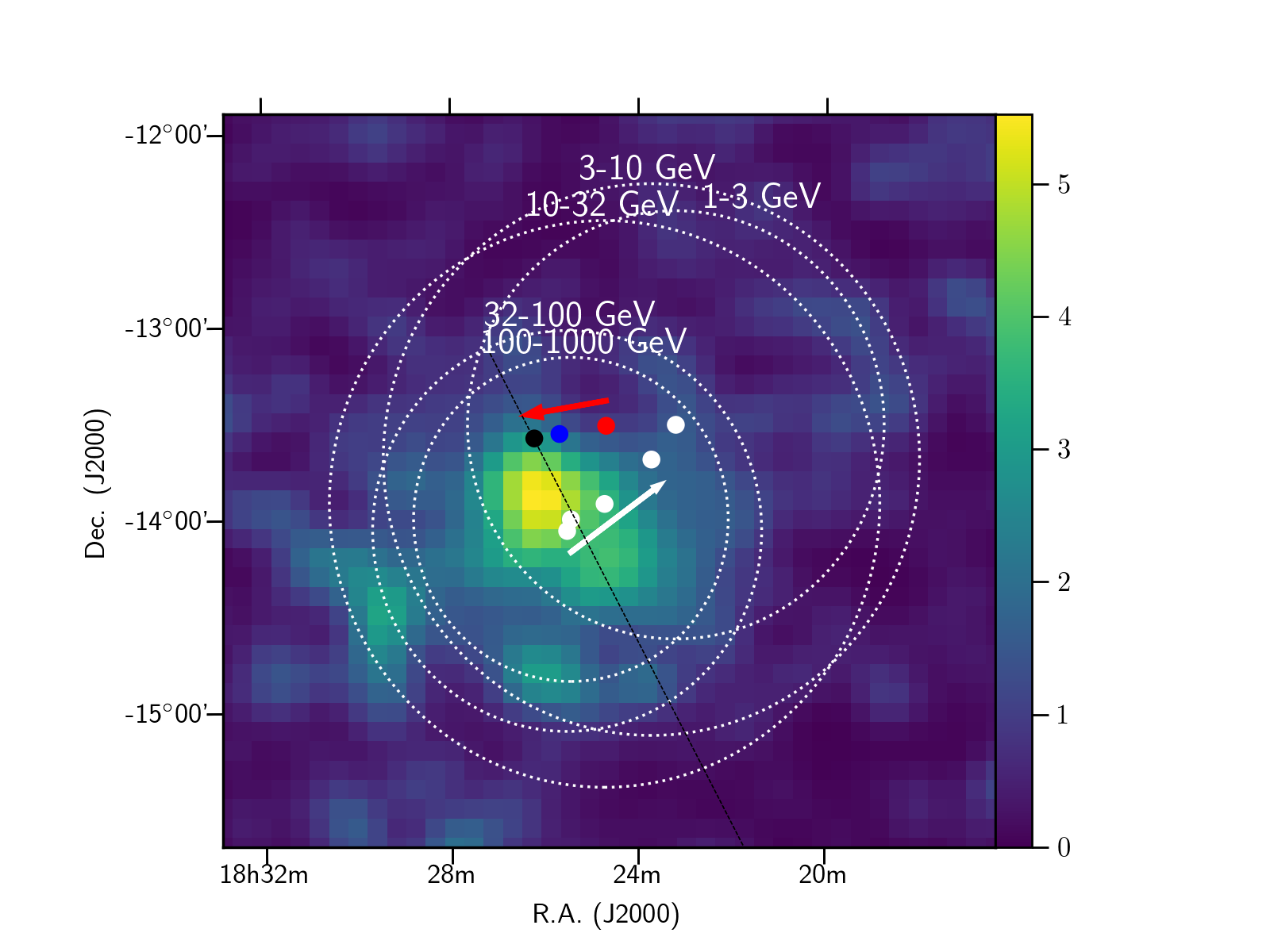}
    \caption{Evolution of the system varying the energy (age) of the emitting particle. Black point: the current PSR position; blue point: birth PSR position at a characteristic age of 21\,kyr; red point: birth PSR position for an hypothetical characteristic age of 60\,kyr. The white points (circles) corresponds to the centroids (extensions) of the PWN estimated using the 2D-Gaussian method in different energy bands (see Table \ref{table_extent}). For the plot, we use the TS map of the energy range 30--100\,GeV as background colour map. The current day major axis for the emission as determined from H.E.S.S. is indicated by a black dashed line.}
    \label{fig:system_evolution}
\end{figure}

In this work, we extended the observations down to 1\,GeV confirming the emission at large distances by the lowest energetic particles.
Figure \ref{fig:system_evolution} shows the size of the PWN for different energy bands. The centroid of the PWN moves with energy and, at high energies (above 30\,GeV), it lies on the H.E.S.S. major emission axis.
The centre position of the PWN seems to move from low to high energy emission, in a direction similar to the pulsar proper motion, which transverse velocity is $v_{\bot}$ = 440 km s$^{-1}$ \citep{Pavlov_2008} (see arrows in Fig. \ref{fig:system_evolution}).
The larger distance between the centroid of the PWN above 100\,GeV (youngest electron population) and below 10\,GeV (oldest electron population) with respect to the expected change in pulsar position ($\sim 0.14^\circ$) due to the proper motion of the pulsar over the estimated characteristic spin-down age, could indicate an higher age for the source or, alternatively, a different preferred direction for the nebula extension in the past.  
Variation in the preferred emission direction may occur due to complex shock interactions. The nebula may be crushed preferentially in certain regions by the  reverse shock returning from the progenitor supernova towards the centre of the system at different times for different directions, such as from nearby molecular clouds. 

\noindent Whilst in the spherically symmetric advection dominated model a constant nebula size is expected below a cooling break energy (Fig. \ref{fig:gameraRE}), a contribution to the energy dependent morphology below 1\,TeV may be attributable to the pulsar proper motion.

\subsection{Particle transport mechanisms}

There are several mechanisms which could account for the travelling and cooling of the electron population inside the nebula. Three are the possibilities previously discussed in \citet{2006A&A...460..365A}: radiative cooling of electrons while they move away from the pulsar causing energy loss; particle transport mechanisms such as diffusion or advection; and variation in the electron injection spectrum over time.

The energy dependent morphology of the nebula can be reproduced with a multi-zone model for electron injection and evolution. The electron population is injected with the same spectrum at the pulsar and evolved in both time (including cooling) and space, assuming dominant advection with a velocity profile of Eq. \eqref{eq:vprof}. This is consistent with both previous models of the PWN \citep{2011ApJ...742...62V} and recent H.E.S.S. results \citep{2019A&A...621A.116H} as well as this analysis. 
 
However, a constant value for $\beta$, the index of the radial and time dependence of the velocity profile, appears not to be favoured; at low energies the data in Fig. \ref{fig:gameraRE} favours the upper edge ($\beta=0.5$), tending towards the lower edge ($\beta=0.75$) at mid-high energies; implying either a change in velocity profile with time and/or that the description is incomplete. Further effects, such as an additional diffusion component as used by \citet{2011ApJ...742...62V} are not included here and may also need to be incorporated for a fuller treatment of the PWN. Additional diffusive effects on top of the bulk advective motion assumed here are not ruled out.

\subsection{The transitional PWN - TeV halo nature of HESS\,J1825$-$137}
The discovery of extended TeV emission around the Geminga pulsar \citep{2017Sci...358..911A}, whose properties are consistent with free particle propagation in the interstellar medium (ISM), suggests the presence of such halo phenomena in other sources \citep{2017ApJ...843...40A} as well as its presence also at GeV energies \citep{2019PhRvD.100l3015D}.
Following the approach of \citet{2020A&A...636A.113G}, we estimated the energy densities for the PWN HESS\,J1825$-$137, and compared it with the typical energy density of the ISM, $\epsilon_{ISM}=0.1$ eV cm$^{-3}$, in order to determine to a first approximation whether the electrons that are responsible for the $\gamma$-ray emission occupy the relatively unperturbed ISM (TeV halo like, $\epsilon_{e} \lesssim \epsilon_{ISM}$ ), or if they are still contained in a region energetically and dynamically dominated by the pulsar (PWN like, $\epsilon_{e} > \epsilon_{ISM}$).

For the estimation of the electron energy density, we divided the total energy of the electron ($W_e$) derived with the NAIMA package (see Table \ref{tab:Naima_fit_params}) by the volume of the nebula. 
As a first approximation, the nebula has been assumed to be a sphere of a radius of $\sim 1.35^{\circ}$ (see Table \ref{table_loc}), which corresponds to a physical radius of 91 pc (2.8$\times10^{20}$ cm) and volume $V \sim 0.92 \times 10^{62}$ cm$^3$. 
Using this basic assumption the mean energy density obtained is $\epsilon_{e,1} = 0.16$ eV cm$^{-3}$.
Taking into account the variability of the extent measurement versus energy, where low energetic particles are living in a wider space than high energetic particles that are more concentrated around the pulsar, we computed the gamma-ray intensity weighted mean of the volume using the full nebula spectrum obtaining a volume of $V \sim 0.84 \times 10^{62}$ cm$^3$. For this case the resulting electron energy density is $\epsilon_{e,2}= 0.17$ eV cm$^{-3}$.
Both the obtained values for the electron energy density ($\epsilon_{e,1}$,$\epsilon_{e,2}$) are compatible to the value derived in \cite{2019arXiv190712121G}, $\epsilon_{e} = 0.25$ eV cm$^{-3}$, supporting the transitional scenario, PWN -- TeV halo like, for this source.  At this stage, high-energy electrons start to escape from the PWN, and propagate into the surrounding supernova remnant, with further escape into the surrounding ISM becoming possible.

\section{Conclusion}
\label{sec:conclude}

Thanks to the first energy-dependent analysis of the HESS\,J1825$-$137 in the GeV range, we found continued morphological changes and increasing size of this PWN towards lower energies. The PWN extent was measured using two complementary approaches; a 2D-Gaussian template fit and the radial profile method as adopted by \citet{2019A&A...621A.116H}, with compatible results. Not only does the PWN extension continue increasing towards lower energies, indicating a possible turnover only below few GeV, but also the best fit centroid of the emission shifts in a direction opposite to the pulsar proper motion. 
Furthermore, considering that the change in the centroid position is larger than the change in pulsar position (see Fig. \ref{fig:system_evolution}), this may be an indication that the system age is somewhat older than that suggested by the 21\,kyr characteristic age of PSR\,J1826$-$1334 or, alternatively, that the preferred direction for the particle transport and therefore nebula extension has varied over time. 
The combined \textit{Fermi}-LAT and H.E.S.S. SED of the nebula could be described by a single electron population model with a break at few hundred GeV. To simultaneously describe the SED and energy dependent morphology of the nebula, we used a multi-zone modelling approach with burst-like injection and an advective velocity profile is found to be consistent with the data. The order $\beta$ of the radial and temporal velocity dependence must, however, vary within the range [0.5,0.75]; a constant velocity profile is not compatible.
The estimated values of the electron energy density support the transitional scenario, PWN – TeV halo like, for this source.

HESS\,J1825$-$137 is one of the most $\gamma$-ray luminous and TeV efficient PWN known \citep{2018A&A...612A...2H}, enabling rich and detailed analyses to be performed. Future studies and modelling of this source as well as similar systems will enable further insights into pulsar wind nebula formation and evolution to be gained.

\subsection*{ACKNOWLEDGMENTS}
The effort of the LAT-team Calibration \& Analysis Working Group to develop Pass 8, as well as the LAT Galactic Working Group for the support given to this project, are gratefully acknowledged.

The Fermi LAT Collaboration acknowledges generous ongoing support from a number of agencies and institutes that have supported both the development and the operation of the LAT as well as scientific data analysis. These include the National Aeronautics and Space Administration and the Department of Energy in the United States, the Commissariat à l'Energie Atomique and the Centre National de la Recherche Scientifique / Institut National de Physique Nucléaire et de Physique des Particules in France, the Agenzia Spaziale Italiana and the Istituto Nazionale di Fisica Nucleare in Italy, the Ministry of Education, Culture, Sports, Science and Technology (MEXT), High Energy Accelerator Research Organization (KEK) and Japan Aerospace Exploration Agency (JAXA) in Japan, and the K. A. Wallenberg Foundation, the Swedish Research Council and the Swedish National Space Board in Sweden.

Additional support for science analysis during the operations phase is gratefully acknowledged from the Istituto Nazionale di Astrofisica in Italy and the Centre National d'Etudes Spatiales in France. This work is performed in part under DOE Contract DE-AC02-76SF00515.

\bibliography{hess1825_papers}  

\begin{appendix}
\onecolumn
\section{Appendix}
\lb{sec:appendix}

\subsection{Visual comparison of the 2D-Gaussian and radial profile extent estimates}
In this section we report the excess maps wth the comparison between the extent obtained using the 2D-Gaussian (see Sect. \ref{ext_2DGaussian}) and the radial profile method (see Sect.\ref{radial_profile_analysis}). 

We also report in Tab. \ref{tab:poly_fit_params} the fit parameters for the polynomial parameterisation in equation \eqref{eq:pol_fit}. In particular, $r_{0}$ is the distance at which it approaches the constant value $c$, and $a$ provides the overall normalisation.

\begin{table}[h]
\centering
\begin{tabular}{c|c|c|c|c}
 Energy (GeV) & $a$ (deg$^{-n}$) & $r_0$ (deg) & $c$  & $\chi^2/$ndf\\
 \hline
 \hline
1 -- 3  & $ 9 \pm 9 $ & $ 3.74 \pm 0.04 $ & $ 68 \pm 60 $ & 54.2 \\
3 -- 10 & $ 1.06 \pm 0.09 $ & $ 3.91 \pm 0.04 $ & $3.9 \pm 0.9$ & 11.9 \\
10 -- 32 & $ (8.62 \pm 0.04) \times 10^{-2}$ & $ 4.30 \pm 0.03 $ & $0.32 \pm 0.02$ & 1.9 \\
32 -- 100 & $(8.19 \pm 0.06)\times 10^{-2}$ & $3.42 \pm 0.03$ & $(6.6 \pm 1.5)\times 10^{-3}$ & 1.4 \\ 
100 -- 1000 & $ (9.2 \pm 0.2) \times 10^{-2} $ & $2.96 \pm 0.05$ & $(0.0 \pm 0.05) \times 10^{-2}$ & 2.5 \\
1 -- 1000 & $ 23 \pm 35 $ & $3.35 \pm 0.02$ & $77 \pm 36$ & 124
\end{tabular}
\caption{Best-fit parameters $a$, $r_0$ and $c$ for the polynomial function equation \eqref{eq:pol_fit} used to parameterise the radial profile in different energy bands, as shown in Fig. \ref{fig:ebandprofsa} and Fig. \ref{fig:ebandprofsb}.}
\label{tab:poly_fit_params}
\end{table}
        
\begin{figure*}[h]
\begin{minipage}{.5\textwidth}
\includegraphics[width=0.97\columnwidth,  trim=15mm 20mm 10mm 20mm,clip=true]{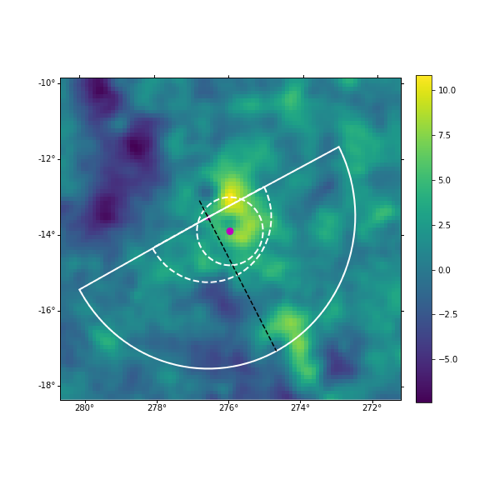}
\end{minipage}
\begin{minipage}{.5\textwidth}
\includegraphics[width=0.97\columnwidth,    trim=0mm 3mm 0mm 3mm,clip=true]{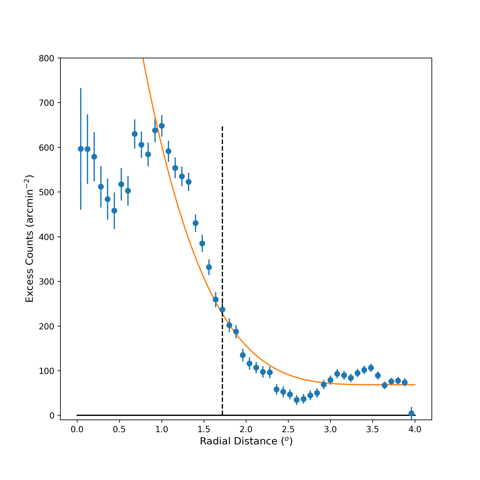}
\end{minipage}%
\\
\begin{minipage}{.5\textwidth}
\includegraphics[width=0.97\columnwidth,    trim=15mm 20mm 10mm 20mm,clip=true]{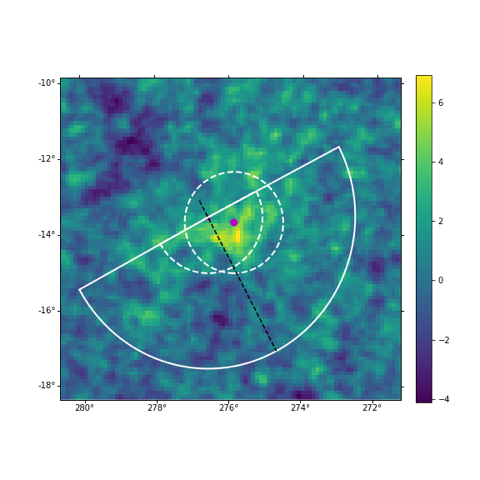}
\end{minipage}%
\begin{minipage}{.5\textwidth}
\includegraphics[width=0.97\columnwidth,    trim=0mm 3mm 0mm 3mm,clip=true]{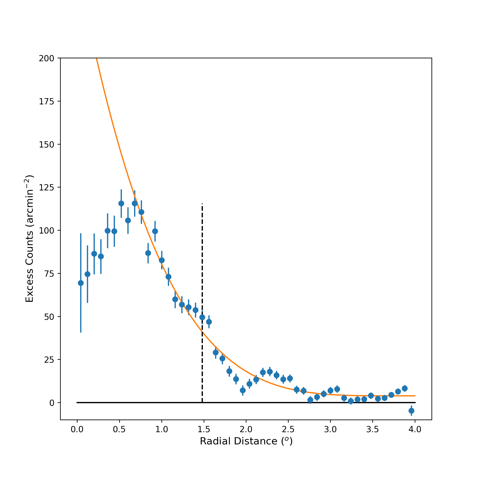}
\end{minipage}%
\caption{\small \label{fig:excess_maps1} Left: Excess maps (in sigma units), in celestial coordinates, of the region around HESS\,J1825$-$137 
The magenta points and the white dashed circles represent the 2D-Gaussian centroid and extension respectively, while the dashed semicircumference shows the extension obtained with the radial profile method considering only the southern hemisphere.
For comparison, the region used to extract the radial profile (as used by H.E.S.S.)is overlaid in white. The preferred emission direction (major axis) as found by H.E.S.S., along which the extent is evaluated, is indicated by the black dashed line. Right: radial profile of the excess counts fit with Eq.\eqref{eq:pol_fit} beyond the peak emission. The characteristic R(1/e) size of the nebula is indicated by a white (black) dashed line in the left (right) hand plot.
The plots are related to the energy bands: 1--3\,GeV (top) and 3--10\,GeV (bottom), }
\label{fig:ebandprofsa}
\end{figure*}

\begin{figure*}[h]
\begin{minipage}{.5\textwidth}
\includegraphics[width=0.97\columnwidth,  trim=15mm 20mm 10mm 20mm,clip=true]{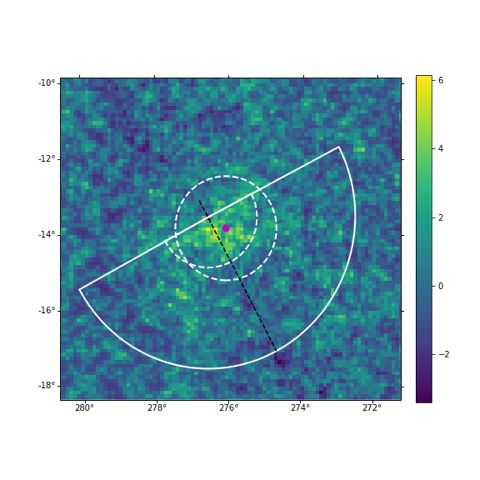}
\end{minipage}
\begin{minipage}{.5\textwidth}
\includegraphics[width=0.97\columnwidth,    trim=0mm 3mm 0mm 3mm,clip=true]{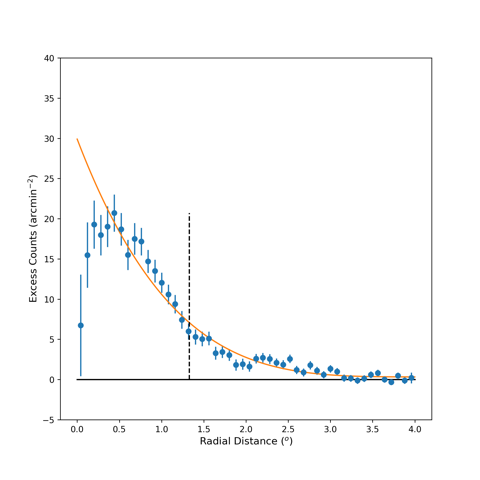}
\end{minipage}%
\\
\begin{minipage}{.5\textwidth}
\includegraphics[width=0.97\columnwidth,    trim=15mm 20mm 10mm 20mm,clip=true]{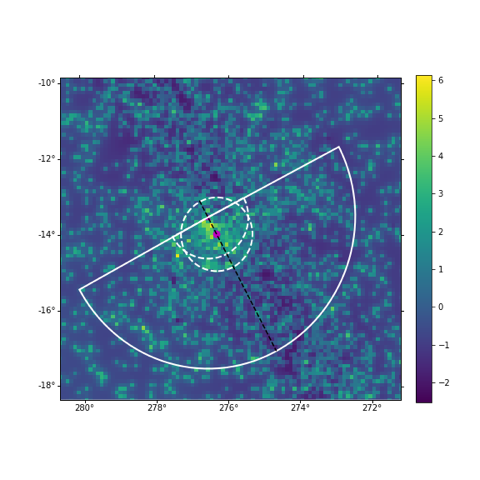}
\end{minipage}%
\begin{minipage}{.5\textwidth}
\includegraphics[width=0.97\columnwidth,    trim=0mm 3mm 0mm 3mm,clip=true]{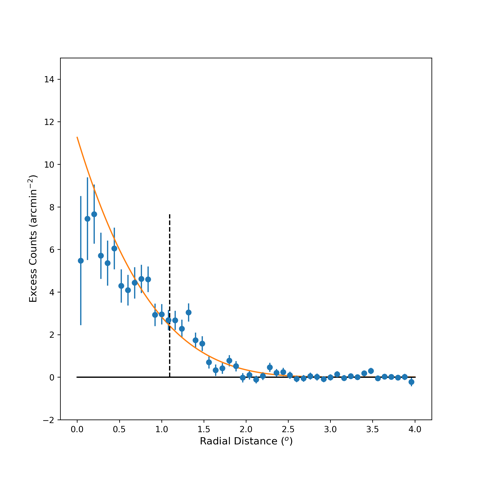}
\end{minipage}%
\\
\begin{minipage}{.5\textwidth}
\includegraphics[width=0.97\columnwidth,    trim=15mm 20mm 10mm 20mm,clip=true]{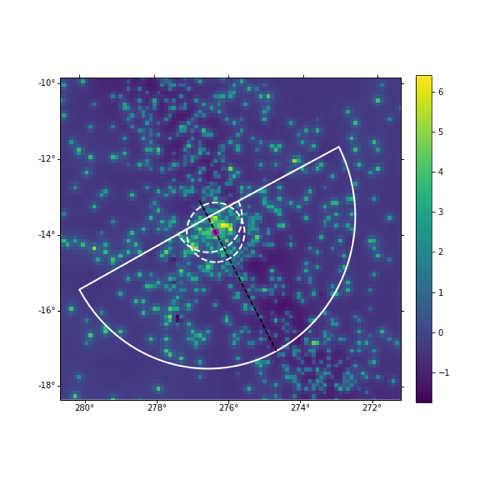}
\end{minipage}%
\begin{minipage}{.5\textwidth}
\includegraphics[width=0.97\columnwidth,    trim=0mm 3mm 0mm 3mm,clip=true]{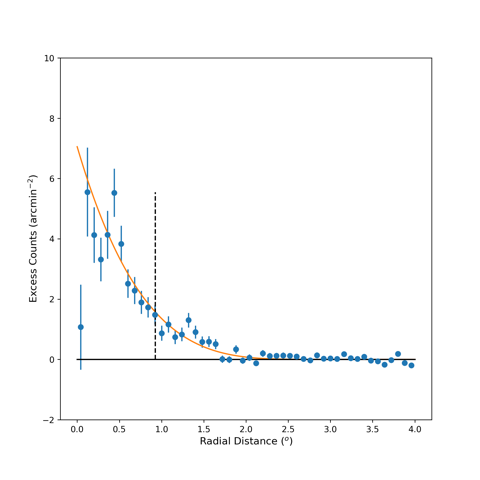}
\end{minipage}%
\caption{\small \label{fig:excess_maps2} 
The labels are the same
as in Fig.\ref{fig:excess_maps1}.
The plots are related to the energy bands: 10--32\,GeV (top), 32--100\,GeV (middle) and 100\,GeV -- 1\,TeV (bottom). }
\label{fig:ebandprofsb}
\end{figure*}

\end{appendix}

\end{document}